\documentclass[aps,superscriptaddress,showpacs,nofootinbib,eqsecnum,prd,notitlepage]{revtex4}

\usepackage{graphics} \usepackage{epsf} \usepackage{eucal}
\usepackage{amssymb} \usepackage{amsmath} \usepackage{graphicx}
\usepackage{bm} \usepackage{dcolumn} \usepackage{epsfig}
\usepackage{multirow} \usepackage{hyperref} \usepackage{subfigure}
\usepackage{appendix} \usepackage{flafter}

\newcommand{\be}{\begin{equation}} \newcommand{\ee}{\end{equation}}
\newcommand{\bea}{\begin{eqnarray}} \newcommand{\eea}{\end{eqnarray}}

\begin{document} 

\title{Cosmological fluctuations of a random field and radiation
  fluid}

\author{Mar Bastero-Gil} \email{mbg@ugr.es} \affiliation{Departamento
  de F\'{\i}sica Te\'orica y del Cosmos, Universidad de Granada,
  Granada-18071, Spain}

\author{Arjun Berera} \email{ab@ph.ed.ac.uk} \affiliation{SUPA, School
  of Physics and Astronomy, University of Edinburgh, Edinburgh, EH9
  3JZ, United Kingdom}

\author{Ian G. Moss} \email{ian.moss@ncl.ac.uk} \affiliation{School of
  Mathematics and Statistics, Newcastlle University, NE1 7RU, United
  Kingdom}

\author{Rudnei O. Ramos} \email{rudnei@uerj.br}
\affiliation{Departamento de F\'{\i}sica Te\'orica, Universidade do
  Estado do Rio de Janeiro, 20550-013 Rio de Janeiro, RJ, Brazil}



\begin{abstract}

A generalization of the random fluid hydrodynamic fluctuation theory
due to Landau and Lifshitz is applied to describe cosmological
fluctuations in systems with radiation and scalar fields. The viscous
pressures, parametrized in terms of the bulk and shear viscosity
coefficients, and the respective random fluctuations in the radiation
fluid are combined with the stochastic and dissipative scalar
evolution equation. This results in a complete set of equations
describing the perturbations in both scalar and radiation
fluids. These derived equations are then studied, as an example, in
the context of warm inflation. Similar treatments can be done for
other cosmological early universe scenarios involving thermal or
statistical fluctuations.

\end{abstract}


\maketitle
\section{introduction}
\label{sec1}

{}Fluctuation and dissipation phenomena could potentially play an
important role in early universe cosmology. When the matter content of
the universe can be split into a subsystem interacting with a large
energy reservoir, then physical processes may be represented through
effective dissipation and stochastic noise terms.  Various physical
systems have been proposed for the early universe which are well
suited for such a treatment.  In particular scenarios where thermal or
any statistical fluctuations seed cosmic structure fit into this
category.  The role of thermal fluctuations in structure formation
have been considered since the early work of Peebles \cite{peeb} and
Harrison \cite{harrison} In recent times, thermal fluctuations
continue to be seen as a possible mechanism for seeding structure
formation
\cite{Moss:1985wn,Berera:1995wh,Berera:1995ie,Berera:1999ws,Magueijo:2002pg,hmb,Bhattacharya:2005wn,Lieu:2011rj,Biswas:2013lna}.
One of the early models with statistical fluctuations seeding
structure formation was in the context of inflationary cosmology with
the warm inflation paradigm \cite{Berera:1995ie}.  There are a variety
of warm inflation models that have been developed (see
\cite{Berera:2008ar,BasteroGil:2009ec} and references therein for
several examples).  Moreover, many models have subsequently been
developed that are very similar to the warm inflation picture, with
particle production during inflation and non-vacuum density
perturbations, such as non-commutative inflation
\cite{Alexander:2001dr}, decaying multifield inflation models
\cite{Battefeld:2008qg}, trapped inflation \cite{Green:2009ds},
cyclic-inflation models \cite{Biswas:2009fv,Biswas:2013lna}, axion
models of inflation \cite{Barnaby:2011qe}, and effective field theory
models of warm inflation \cite{LopezNacir:2011kk}. Thermal
fluctuations have also been examined as the origin of density
perturbations in bouncing universe models
\cite{Magueijo:2002pg,Cai:2009rd}, string cosmology
\cite{Nayeri:2005ck,Brandenberger:2006vv}, loop cosmology
\cite{Magueijo:2007wf}, the near-Milne universe
\cite{Magueijo:2007na}, a model of phase transition involving
holography \cite{Magueijo:2006fu}, and varying speed of light models
\cite{Magueijo:2002pg}.  Density perturbations created from a
statistical state have been examined in the radiation dominated regime
\cite{Magueijo:2002pg,hmb,Bhattacharya:2005wn,Lieu:2011rj,Biswas:2013lna}
and other excited states \cite{Landau:2011aa,Ashoorioon:2013eia}.
During reheating and preheating, particle production has been shown to
affect cosmological perturbations 
\cite{Jain:2009ep,Leung:2012ve,Mukaida:2014yia}.

In all these scenarios a common feature is a sizable
density of particles in the universe that is pictured in some
statistical state, usually thermal.  In order to thermalize, these
particles must interact with one another.  The short scale physics in
the early universe is not directly accessible to observation today.
Moreover this is a complicated many-body problem, which, similar to
such problems in condensed matter systems, would be effectively
intractable to exact calculation.  The typical approach to such a
problem is for this many-body dynamics to be characterized by one,
perhaps many, microphysical scales, within which one can employ a
statistical treatment and the dynamics manifests itself through
dissipation processes.  Associated with such effects will be
corresponding stochastic forces.

A treatment involving fluctuation-dissipation dynamics can be
implemented at different levels of coarse graining of the degrees of
freedom.  Ideally one should start with the underlying fundamental
quantum field theory and proceed to coarse grain.  This level of
sophistication has been realized in some simple cases, most notably
the Caldeirra-Leggett model for condensed matter systems
\cite{clmodel} and this model has also been examined in a cosmological
context \cite{Berera:1996nv}.  In this case the quantum model can be
coarse grained systematically into a stochastic langevin equation for
the system, with the remaining degrees of freedom represented as a
noise force and a dissipative term.  In more complicated quantum field
theory models, coarse graining has been done at a perturbative level
\cite{Hosoya:1983ke,Morikawa:1986rp,GR,BMR,Miyamoto:2013gna}.  Such
approaches treat one part of the quantum field theory as a system and
then integrate out the remaining fields into a heat reservoir.

In treating cosmological perturbations, the problem is a little more
involved than simply deriving the stochastic evolution equation for
the system.  The heat bath will also have fluctuation-dissipation
effects due to the action of the system as well as internal effects
from within the heat bath.  These effects will play a role in the
cosmological perturbation equations and have consequences for the
density perturbations.  This is a much harder problem.  Deriving the
full system-heat bath dynamics from quantum field theory with all
fluctuation and dissipation effects computed to our knowledge has
never been achieved.  An intermediate approach is to treat the heat
bath within a fluid approximation which is then coupled to the system,
which is treated from quantum field theory.  It is at this level that
the study in this paper proceeds.  In this approximation, the heat
bath is treated in terms of quantum fields for computing the transport
and noise coefficients of the system and the heat bath itself.
However, in treating the cosmological perturbations, the heat bath is
then represented as a fluid.  The missing step here is showing how the
quantum fields that comprise the heat bath can be represented as a
fluid.  At an intuitive level the correspondence seems evident, but it
is a difficult problem of coarse graining that goes well beyond the
concerns of the specific problem being addressed in this paper.  This
limitation in our approach thus brings some lack of precision in
formulating the dynamical problem. Nevertheless, it still captures a
great deal of the physics that otherwise is completely ignored in
simple mean field treatments.

Once the problem is formulated in this way, progress can be made.
Hydrodynamics is a macroscopic theory describing the behavior of
averaged or mean variables corresponding to the energy density,
pressure, fluid velocity and so on. As such, the microscopic physics
become manifest in the form of dissipative terms corresponding to the
transport coefficients, like bulk and shear viscosities.  But from a
fluctuation-dissipation stand point, these must also be related to
stochastic fluctuations as well. Landau and Lifshitz were the first to
propose a fluctuating hydrodynamics theory, where random fluxes are
added to the usual hydrodynamic equations, with two-point correlation
functions related with the transport coefficients through
fluctuation-dissipation relations~\cite{landau}. The Landau-Lifshitz
fluctuating hydrodynamics theory was later refined and put in firm
theoretical grounds by the work of Fox and Uhlenbeck~\cite{fox} and
extended to the relativistic fluids by Zimdahl~\cite{zimdahl1} (see
also \cite{calzetta}).

Aside for warm inflation, most cosmological models involving thermal
or statistical fluctuations have been examined at only a mean field
level, where fluctuation-dissipation effects have not been treated.
As such, important information is ignored about how the short scale
physics affects the large scale physics.  Calculations in warm
inflation have shown that current precision from CMB data demands a
treatment beyond a mean field level, and requires account for
fluctuation and dissipation effects.  This lesson probably also
carries over for other scenarios involving thermal fluctuations.

In this paper, we will study the density perturbations spectra in
terms of the coupled set of radiation equations describing the random
radiation fluid equations and the stochastic equation for a scalar
field.  We believe this is the first study of such a system.  The
study of cosmological perturbations making use of the relativistic
version of the fluctuation hydrodynamics theory of Landau and Lifshitz
has already been done before by Zimdahl in \cite{zimdahl2}.  Some
papers have treated the density perturbations in a system of a scalar
field with dissipation coupled to a radiation fluid \cite{hwangnoh} as
well as affects of viscosity within the radiation fluid
\cite{shear,bulkback}.  However no work has treated in addition the
corresponding noise forces that accompany dissipation and viscosity.
Our treatment in this paper includes all these effects and can be
applied to problems in cosmology involving a scalar field coupled to
other systems, such as in inflationary cosmology, cosmic phase
transitions, reheating, curvaton decay etc...  Often, in addition of
including the scalar field dynamics, we also have a mechanism by which
the radiation bath is generated and maintained through particle
production due to the decay of the scalar field.  We analyze in detail
not only the interplay of the different dissipation terms, from the
scalar field and the bulk and shear viscosities of the radiation
fluid, but also the effect of the respective noise terms, connected
with the dissipation and viscosity terms by the
dissipation-fluctuation relations.

This paper is organized as follows. In Sec.~\ref{sec2} we introduce
the relativistic fluctuating hydrodynamics built from the original
version due to Landau and Lifshitz.  In Sec.~\ref{sec3}, guided by the
equivalence principle, we extend the relativistic fluctuating
hydrodynamic equations for the radiation bath for the cosmological
context. The equations are coupled with those of the inflaton field as
appropriate in dissipative environments, like in the warm inflation
scenario, and the perturbation equations constructed.  In
Sec.~\ref{diss+visc} we give the general expressions for the
dissipative and viscosity coefficients we will be considering along
this work.  In Sec.~\ref{sec5}, the full cosmological perturbations
are studied numerically and results for the curvature power spectrum
for perturbations presented. The effects of both bulk and shear
viscosities are analyzed.  {}Finally, in Sec. \ref{sec6} we give our
concluding remarks.

\section{Fluctuations in flat spacetime}
\label{sec2}

We are interested primarily in situations with a radiation fluid that
is close to being in thermal equilibrium at some local temperature
$T$, and the fluid is hot enough to be treated as classical and
relativistic.  Quantum statistical mechanical fluctuations in such a
radiation fluid can be described using Landau's theory of random
fluids \cite{landau}, where the deterministic equations of fluid
dynamics are replaced by a system of equations with stochastic source
terms.  The fluid approximation is maintained by microscopic
interactions, with small departures from equilibrium which cause both
fluctuations and dissipation. The fluctuations of the fluid reach a
balance between the effects of the source and the dissipation
terms. Fixing the statistical properties of the noise terms to ensure
that stochastic averages of fluid variables reproduce the statistical
ensemble averages leads to the fundamental fluctuation-dissipation
relation.

In this work we will be using the first-order (or Eckart) dissipative relativistic 
hydrodynamics. Even though the first-order formalism is known to have problems 
concerning causality, stability and in general they do not have
a well-posed initial value formulation, the first order formalism of hydrodynamics is 
still simpler and more immediate to study in the context of 
cosmological perturbations than the second and higher order formalisms of hydrodynamics
(for a recent review, see, e.g., ref.~\cite{roma} and also ref.~\cite{kumar}
for the case of including fluctuations).  {}Furthermore, 
the use of the Eckart first order formalism can be justified on general grounds 
for small enough radiation bath relaxation times~\cite{bulkback},
which is one of the conditions required to justify a 
close to equilibrium thermal bath, and assumed in the application 
performed in section~\ref{sec5}.

Consider a relativistic fluid with energy density $\rho^{(f)}$ and
pressure $p^{(f)}$ in which conserved particle numbers are absent or
negligible, and the 4-velocity $u^{a(f)}$ is the velocity of energy
transport. Random sources and dissipative stresses are introduced via
a stress term $\Pi_{ab}$ in the stress-energy tensor,

\begin{equation}
T^{(f)}{}_{ab}=(p^{(f)}+\rho^{(f)})\,u_a{}^{(f)} u_b{}^{(f)}+p^{(f)}
\,g_{ab}+\Pi_{ab},\label{fset}
\end{equation}
where indices $a,b\dots$ denote spacetime components.  In Landau's
theory, dissipation is governed by constitutive relations for shear
viscosity $\eta_s$ and bulk viscosity $\eta_b$ whilst fluctuations are
generated by a Gaussian noise term $\Sigma_{ab}$. In a comoving frame
where the spatial components $u_i{}^{(f)}=0$ and the time component
$u_0{}^{(f)}=-1$, the non-vanishing shear terms are

\begin{equation}
\Pi_{ij}=-\left(\eta_s\nabla_i u_j{}^{(f)}+\eta_s\nabla_j u_i{}^{(f)}
+(\eta_b-\frac23\eta_s)\delta_{ij} \nabla_k u^{(f)k}\right)
-\Sigma_{ij},\label{sst0}
\end{equation}
where $\nabla_i$ denotes a spatial derivative.  The correlation
functions of the stochastic noise term $\Sigma_{ij}$ are assumed to be
local and determined by the fluctuation-dissipation relation,

\begin{equation}
\langle\Sigma_{ij}(x,t)\Sigma_{kl}(x',t')\rangle=
2T\left(\eta_s\delta_{ik}\delta_{jl}+\eta_s\delta_{il}\delta_{jk}+
(\eta_b-\frac23\eta_s)\delta_{ij}\delta_{kl}\right)
\,\delta^{(3)}(x-x')\delta(t-t').\label{ncr}
\end{equation}
This will be explored further in Sect. \ref{fdt}.  Landau's theory can
be used reliably for small departures from a stable underlying fluid
flow. We shall be concerned mostly with small density, pressure
and 3-velocity fluctuations
$\delta{\bf u}^{(f)}$ of a radiation fluid in an inertial frame with
background density $\rho^{(f)}$ and pressure $p^{(f)}$.
For example,
the momentum conservation equation obtained using the vanishing
divergence of the stress-energy tensor outlined above is,

\begin{equation}
(p^{(f)}+\rho^{(f)})\delta\dot{\bf u}^{(f)}+\boldsymbol{\nabla}\delta
  p +\dot p^{(f)}\,\delta{\bf u}^{(f)}
  =\eta_s\boldsymbol{\nabla}^2\delta{\bf u}^{(f)}
  +\frac13\eta_s\boldsymbol{\nabla}\nabla\cdot\delta{\bf u}^{(f)}
  +\eta_b \boldsymbol{\nabla}\nabla\cdot\delta{\bf u}^{(f)}
  +\boldsymbol{\nabla}\cdot\boldsymbol{\Sigma}.\label{fff}
\end{equation}
This can be recognized as the perturbed Navier-Stokes momentum
conservation equation with a stochastic source term. 
The solutions to the stochastic fluid equations can be used to follow
the evolution of quantities such as the density perturbations,
\begin{equation}
\langle\delta\rho^{(f)}(x,t)\delta\rho^{(f)}(x',t)\rangle,
\end{equation}
by taking a stochastic average. Without the theory of random fluids,
we would only have knowledge of the equilibrium values of the density
fluctuations.
\subsection{Relativistic fluids coupled to a scalar field}

Our aim is to couple this radiation fluid to a scalar field. The
behavior of a relativistic scalar field in flat spacetime interacting
with radiation can be analyzed using non-equilibrium quantum field
theory \cite{schwingerkeldysh}.  When the small-scale behavior of the
fields is averaged out, the scalar field fluctuations, like the fluid
fluctuations, can be described by a stochastic system whose evolution
is determined by a Langevin equation \cite{Hu-book}.  For a weakly
interacting radiation gas, the dissipation and noise terms in the
Langevin equation can be approximated by local expressions. This is
the case we will consider here. The Langevin equation for a scalar
field with thermodynamic potential $\Omega(\phi,T)$ and damping
coefficient $\Upsilon(\phi,T)$ is then~\cite{GR}

\begin{equation}
-\Box\phi(x,t)+\Upsilon\dot\phi(x,t)+\Omega_{,\phi} =(2\Upsilon
T)^{1/2}\xi^{(\phi)}(x,t),\label{stoc}
\end{equation}
where $\Box$ is the flat spacetime d'Alembertian and $\xi^{(\phi)}$ is
a stochastic source. The probability distribution of the source term
will be approximated by a localized gaussian distribution with
correlation function \cite{GR,BMR},

\begin{equation}
\langle\xi^{(\phi)}(x,t)\xi^{(\phi)}(x',t')\rangle=\delta^{(3)}(x-x')\delta(t-t').\label{phin}
\end{equation}
The Langevin equation applies when the surrounding radiation is at
rest.  For a fluid in uniform motion with 4-velocity $u^{a(f)}$ we
would need to choose the Lorentz frame to be the rest frame of the
fluid. This can be expressed in covariant form by replacing $\dot\phi$
in the dissipation term by a fluid derivative,

\begin{equation}
D\phi=u^{a(f)}\nabla_a\phi.\label{der}
\end{equation} 
The dissipation results in a transfer of energy and momentum from the
scalar field to the radiation which needs to be included in the fluid
equations.

Energy and momentum transfer can be tracked by considering the
divergence of the stress-energy tensor. We combine the fluid and
scalar contributions into a unified stress-energy tensor given by

\begin{equation}
T_{ab}=Ts\,u^{(f)}_a u^{(f)}_b-\Omega \,g_{ab}+
\nabla_a\phi\nabla_b\phi-\frac12(\nabla\phi)^2g_{ab}+\Pi_{ab},\label{set}
\end{equation}
where $\Pi_{ab}$ is orthogonal to the fluid velocity.  We have
introduced the entropy density $s$, defined by the thermodynamic
relation

\begin{equation}
s=-\Omega_{,T}.
\end{equation}
If $s_{,\phi}\equiv0$, then the thermodynamic potential $\Omega$
splits into an effective potential $V(\phi)$ depending only on $\phi$
and a radiation term depending only on $T$,

\begin{equation}
\Omega=V(\phi)-p^{(f)}(T).
\end{equation}
In this case, $s\equiv s(T)$ and the fundamental thermodynamic
relation implies that $Ts=\rho^{(f)}+p^{(f)}$ allowing us to separate
off the fluid stress-energy tensor $T^{(f)}{}_{ab}$ given in
Eq. (\ref{fset}). This separation into fluid and scalar field terms is
not possible in general, but a partial separation can be seen in the
divergence of the stress-energy tensor,

\begin{equation}
\nabla^bT_{ab}=\left(D(Ts)+\nabla_bu^{b(f)}\right)u_a{}^{(f)}{}+s\nabla_aT
+\nabla^b\Pi_{ab} +\left(\Box\phi-\Omega_{,\phi}\right)\nabla_a\phi.
\end{equation}
The first three terms represent the field equations for the fluid in
the absence of the scalar field and they can be separated from the
remaining terms by defining fluxes $Q^{(f)}_a$ and $Q^{(\phi)}_a$ by,

\begin{eqnarray}
Q^{(f)}_a&=&\left(D(Ts)+\nabla_bu^{(f)b}\right)u^{(f)}{}_a+s\nabla_aT
+\nabla^b\Pi_{ab},\\ Q^{(\phi)}_a&=&\left(\Box\phi-\Omega_{,\phi}\right)\nabla_a\phi.
\end{eqnarray}
Using the Langevin Eq. (\ref{stoc}) for the scalar field, we obtain
that

\begin{equation}
Q^{(\phi)}_a= \Upsilon(D\phi)\nabla_a\phi-(2\Upsilon
T)^{1/2}\xi^{(\phi)}\nabla_a\phi.
\label{flux}
\end{equation}
Energy-momentum conservation $\nabla^bT_{ab}=0$ results in a set of
fluid equations,

\begin{equation}
\left(D(Ts)+\nabla_bu^{(f)b}\right)u^{(f)}{}_a+s\nabla_aT
+\nabla^b\Pi_{ab}=-Q^{(\phi)}_a.
\end{equation}
Therefore, the flux $Q^{(\phi)}_a$ describes the transfer of energy
and momentum to the fluid equations.

As a matter of fact, Eq. (\ref{flux}) is not the most general
expression which we can obtain for the energy transfer. We might also
consider adding a stochastic energy flux term $\boldsymbol{P}$ to the
stress energy tensor, rather like the stochastic stress term
$\Sigma_{ij}$ which we had in Eq. \ref{sst0}, so that the stress
energy tensor becomes
\begin{equation}
T_{ab}=Ts\,u^{(f)}_a u^{(f)}_b-\Omega \,g_{ab}+
\nabla_a\phi\nabla_b\phi-\frac12(\nabla\phi)^2g_{ab}+\Pi_{ab}+2u_{(a}{}^{(f)}P_{b)}.
\end{equation}
This modifies the energy transfer vector $Q^{a(\phi)}$,

\begin{equation}
Q^{(\phi)}_a= \Upsilon(D\phi)\nabla_a\phi-(2\Upsilon
T)^{1/2}\xi^{(\phi)}\nabla_a\phi
+\nabla^b\left(2u_{(a}{}^{(f)}P_{b)}\right).\label{flux2}
\end{equation}
The time component represents energy transfer,

\begin{equation}
Q_0^{(\phi)}=\Upsilon(D\phi)\dot\phi-(2\Upsilon
T)^{1/2}\xi^{(\phi)}\dot\phi -\boldsymbol{\nabla}\cdot\boldsymbol{P}.
\end{equation}
The simplest possibility is simply $P_a=0$, but an interesting
alternative is to impose the condition that the energy flux is
independent of $\xi^{(\phi)}$, by setting

\begin{equation}
\boldsymbol{\nabla}\cdot\boldsymbol{P}=-(2\Upsilon
T)^{1/2}\xi^{(\phi)}\dot\phi.
\end{equation}
In this case $\boldsymbol{P}$ has to be included in the momentum flux
$\boldsymbol{Q}^{(\phi)}$.  The calculations in later sections will
consider both of these possibilities.
 
\subsection{Perturbation theory}

We perturb the fluid quantities and the scalar field, replacing
$\rho^{(f)}$ by $\rho^{(f)}+\delta\rho^{(f)}$ and so on, and taking
the backgrounds to be homogeneous with vanishing velocity.  From this
point on we use the indices $i,j\dots$ to denote the spatial
coordinate frame in which the background fluid is at rest.  The
non-vanishing components of the stress tensor $\Pi_{ab}$ are given by
the constitutive relations for shear and bulk viscosity as well as the
random noise term $\Sigma_{ij}$ generating the fluctuations,

\begin{equation}
\Pi_{ij}=-\left(\eta_s\nabla_i\delta u^{(f)}_j+\eta_s\nabla_j\delta
u^{(f)}_i +(\eta_b-\frac23\eta_s)\delta_{ij} \nabla_k\delta
u^{(f)k}\right) -\Sigma_{ij}.\label{sst}
\end{equation}
The noise term is taken to be gaussian with the correlation function
(\ref{ncr}).  The first-order fluid equations obtained from
energy-momentum conservation $\nabla^bT_{ab}=0$ using the
stress-energy tensor (\ref{set}) are then

\begin{eqnarray}
T\,\delta\dot s+\dot s\,\delta T
+Ts\,\boldsymbol{\nabla}\cdot\delta{\bf u}^{(f)}&=& -\delta
Q^{(\phi)},\\ \{Ts\,\delta{\bf
  u}^{(f)}\}\dot{\phantom{a}}+\boldsymbol{\nabla}(s\delta T)
-\eta_s\nabla^2\delta{\bf u}^{(f)}-\left(\eta_b+\frac13\eta_s\right)
\boldsymbol{\nabla}\nabla\cdot\delta{\bf u}^{(f)} &=&-\delta {\bf
  Q}^{(\phi)}+\boldsymbol{\nabla}\cdot\boldsymbol{\Sigma},\label{ff}
\end{eqnarray}
where boldface denotes spatial vectors and $\delta Q^{(\phi)}=\delta
Q^{(\phi)0}=-\delta Q^{(\phi)}_0$.  Comparison with the random fluid
Eq. (\ref{fff}) suggests that we should identify the fluid density and
pressure perturbations as

\begin{eqnarray}
\delta\rho^{(f)}&=&T\,\delta s,\label{drho}\\ \delta
p^{(f)}&=&s\,\delta T.\label{dp}
\end{eqnarray}
The fluctuations $\delta\rho^{(f)}$, $\delta p^{(f)}$ and $\delta\phi$
are obtained from just two thermodynamical degrees of freedom $\phi$
and $T$, so one of the fluctuations is dependent on the other two, the
natural choice being the pressure perturbation. By setting $\delta
s=s_{,\phi}\delta\phi+s_{,T}\delta T$ in (\ref{drho}), we arrive at

\begin{equation}
\delta
p^{(f)}=c_s^2(\delta\rho^{(f)}-Ts_{,\phi}\delta\phi),\label{dpp}
\end{equation}
where the sound speed $c_s^2=s/(Ts_{,T})$. Differentiating Eqs.
(\ref{drho}) and (\ref{dp}), we also have

\begin{equation}
T\delta \dot s+\delta T\dot s= \delta\dot\rho^{(f)}+s_{,\phi}\delta q,
\end{equation}
where we have defined

\begin{equation}
\delta q=\dot\phi\,\delta T-\dot T\,\delta\phi.
\end{equation}
The fluid equations can then be re-written in terms of the density and
scalar field fluctuations,

\begin{eqnarray}
\delta\dot\rho^{(f)}
+(\rho^{(f)}+p^{(f)})\,\boldsymbol{\nabla}\cdot\delta{\bf u}^{(f)}
+s_{,\phi}\delta q&=& -\delta
Q^{(\phi)},\label{ffs1}\\ \{(\rho^{(f)}+p^{(f)})\,\delta{\bf
  u}^{(f)}\}\dot{\phantom{a}}+\boldsymbol{\nabla}\delta p^{(f)}
-\eta_s\nabla^2\delta{\bf u}^{(f)}-\left(\eta_b+\frac13\eta_s\right)
\boldsymbol{\nabla}\nabla\cdot\delta{\bf u}^{(f)} &=&-\delta {\bf
  Q}^{(\phi)}
+\boldsymbol{\nabla}\cdot\boldsymbol{\Sigma}.\label{ffs2}
\end{eqnarray}
When $s_{,\phi}\equiv0$, then $\delta p^{(f)}=c_s^2\delta\rho^{(f)}$
and the $\delta q$ term drops out of the fluid equations. In this case
the equations become perturbed versions of the relativistic
Navier-Stokes equations with stochastic source terms.

Since there are no sources of vorticity at linear order, we can
introduce scalar velocity perturbations through

\begin{equation}
\delta{\bf u}^{(f)}=\boldsymbol{\nabla}\delta v^{(f)},\qquad
\delta{\bf Q}^{(\phi)}=\boldsymbol{\nabla}\delta J^{(\phi)}.
\end{equation}
The fluid perturbations for potential flow satisfy

\begin{eqnarray}
\delta\dot\rho^{(f)} +(\rho^{(f)}+p^{(f)})\,\nabla^2\delta v^{(f)}
+s_{,\phi}\delta q&=& -\delta
Q^{(\phi)},\label{fsf1}\\ \{(\rho^{(f)}+p^{(f)})\,\delta
v^{(f)}\}\dot{\phantom{a}}+\delta p^{(f)}-\eta^\prime\nabla^2\delta
v^{(f)} &=&-\delta J^{(\phi)}+(2\eta^\prime
T)^{1/2}\xi^{(f)},\label{fsf2}
\end{eqnarray}
where $\delta p^{(f)}$is given by Eq. (\ref{dpp}) and we have defined
$\eta^\prime$ as the combination of viscosity coefficients:

\begin{equation}
\eta^\prime=\frac43\eta_s+\eta_b.
\end{equation}
Using Eq. (\ref{ncr}), the noise source
$\xi^{(f)}=\nabla^{-2}\nabla^i\nabla^j\Sigma_{ij}$ has correlation
function

\begin{equation}
\langle\xi^{(f)}(x,t)\xi^{(f)}(x',t')\rangle=\delta^{(3)}(x-x')\delta(t-t').\label{fnoise}
\end{equation}
The new feature of these equations is that they combine the random
fluid with the exchange of energy and momentum to the scalar field,
represented by the flux terms $\delta Q^{(\phi)}$ and $\delta {\bf
  Q}^{(\phi)}$.  {}For a homogeneous background scalar field, the
perturbation of Eq. (\ref{flux2}) shows that

\begin{eqnarray}
\delta
Q^{(\phi)}&=&-\delta\Upsilon\dot\phi^2-2\Upsilon\dot\phi\,\delta\dot\phi
+(2\Upsilon
T)^{1/2}\dot\phi\,\xi^{(\phi)}+\boldsymbol{\nabla}\cdot\boldsymbol{P}
\label{fsq}\\
\delta J^{(\phi)}&=&\Upsilon\dot\phi\,\delta\phi
+\nabla^{-2}\boldsymbol{\nabla}\cdot\boldsymbol{\dot P}.\label{fsj}
\end{eqnarray}
We shall take

\begin{equation}
\boldsymbol{P}=-C_P(2\Upsilon
T)^{1/2}\dot\phi\,\nabla^{-2}\boldsymbol{\nabla}\xi^{(\phi)}.
\end{equation}
The two cases $C_P=0$ and $C_P=1$ govern whether the noise source
$\xi^{(\phi)}$ appears in the energy flux or in the momentum
flux. Both cases will be considered in our numerical analysis to be
performed in Sec.~\ref{sec5}.
The procedure here assumes
a linear transfer of energy from the $\phi$-system to the radiation fluid,
so that the $\phi$ noise term at some mode ${\bf k}$ 
transfers energy into mode ${\bf k}$
of the radiation fluid.  In a quantum field theory the radiation
fluid would be associated with the effective quadratic parts
of the light fields in the system.
In general there will be nonlinear terms
transferring energy between the $\phi$-field and this fluid.  
Thus, in associating
the hydrodynamic approximation developed in this paper to an underlying
quantum field theory system, this possibility of nonlinear couplings must
be considered, though we will not develop this point any further in this
paper.

\subsection{Fluctuation-dissipation relations}\label{fdt}

We finish this section with a discussion of the
fluctuation-dissipation relations to verify that the stochastic
average

\begin{equation}
\langle\delta\rho^{(f)}(x,t)\delta\rho^{(f)}(x',t)\rangle,
\end{equation}
reproduces the quantum-statistical ensemble average on
time-independent backgrounds.  This is expected on general grounds,
but the derivation for relativistic fields is less well known than the
non-relativistic case and the density correlations will be useful
later. The thermal ensemble averages can be obtained using standard
thermodynamical arguments, or by using thermal quantum field theory
(see \cite{hmb} for an example). These thermodynamic results have also been
used in cosmological settings, e.g. by \cite{hmb,Biswas:2013lna}.

We disconnect the scalar field by setting $Q^{(\phi)}_a=s_{,\phi}=0$
and take the background density and pressure to be constant. This
allows Fourier decomposition with

\begin{eqnarray}
\delta\rho^{(f)}(k,\omega)&=&\int dt\,d^3
x\,\delta\rho^{(f)}(x,t)\,e^{i(k\cdot x-\omega t)},\\ \delta
v^{(f)}(k,\omega)&=&\int dt\,d^3 x\,\delta v^{(f)}(x,t)\,e^{i(k\cdot
  x-\omega t)}.
\end{eqnarray}
On substituting these transforms into Eqs. (\ref{ffs1}) and
(\ref{ffs2}), the fluctuations satisfy

\begin{equation}
\begin{pmatrix}
i\omega&-(1+c_s^2)\rho^{(f)}
k^2\\ c_s^2&i\omega(1+c_s^2)\rho^{(f)}+k^2\eta'
\end{pmatrix}
\begin{pmatrix}
\delta\rho^{(f)}\\ \delta v^{(f)}
\end{pmatrix}
=(2\eta'T)^{1/2}
\begin{pmatrix}
0\\ \xi^{(f)}
\end{pmatrix}.
\end{equation}
The solution for the density fluctuation is

\begin{equation}
\delta\rho^{(f)}(k,\omega)=G(k,\omega)\,k^2(2\eta'T)^{1/2}\xi^{(f)},
\end{equation}
with the Green function

\begin{equation}
G(k,\omega)=\left[(\gamma k^2-i(\omega-c_s k)) (\gamma
  k^2-i(\omega+c_s k))-\gamma^2k^4\right]^{-1},
\end{equation}
and $\gamma=\eta'/2(1+c_s^2)\rho^{(f)}$. After using the noise
correlation function (\ref{fnoise}), the density correlation functions
become

\begin{equation}
\langle\delta\rho^{(f)}(k,t)\delta\rho^{(f)}(k',t)\rangle=
\int{d\omega\over
  2\pi}|G(k,\omega)|^2(2\eta'T)k^4(2\pi)^2\delta^{(3)}(k+k').
\end{equation}
In the low damping regime $\gamma k \ll c_s $, the integration gives
\begin{equation}
\langle\delta\rho^{(f)}(k,t)\delta\rho^{(f)}(k',t)\rangle\approx
                 {1+c_s^2\over
                   c_s^2}\,T\rho^{(f)}(2\pi)^2\delta^{(3)}(k+k').\label{rr1}
\end{equation}
{}For comparison, statistical mechanics relates the fluctuations at
temperatures large enough to ignore quantum effects to the entropy
density $s$ \cite{hmb},

\begin{equation}
\langle\delta\rho^{(f)}(k,t)\delta\rho^{(f)}(k',t)\rangle_{\rm
  sm}\approx T^3{\partial s\over\partial T}(2\pi)^2\delta^{(3)}(k+k').
\end{equation}
In the case where the density depends only on temperature, we have

\begin{equation}
\rho^{(f)}=aT^{1+1/c_s^2},\qquad s=a(1+c_s^2) T^{1/c_s^2} .
\end{equation}
It follows that

\begin{equation}
\langle\delta\rho^{(f)}(k,t)\delta\rho^{(f)}(k',t)\rangle_{\rm
  sm}\approx {1+c_s^2\over
  c_s^2}\,T\rho^{(f)}(2\pi)^2\delta^{(3)}(k+k').\label{rr2}
\end{equation}
Equations (\ref{rr1}) and (\ref{rr2}) agree, confirming that the
coefficient of the noise term was chosen correctly.

The fluctuation-dissipation relations for the scalar field can be
obtained by following a similar route.  We take a constant background
scalar field and consider the fluctuations $\delta\phi$. Their
{}Fourier transforms satisfy

\begin{equation}
\delta\phi(k,\omega)=G(k,\omega)\,(2\Upsilon T)^{1/2}\xi^{(\phi)},
\end{equation}
where the Green function is

\begin{equation}
G(k,\omega)=(k^2-\omega^2-i\Upsilon\omega+m^2)^{-1},
\end{equation}
and $m^2=V_{,\phi\phi}$. {}Following the same steps as above, with
$\Upsilon \ll k$, these give

\begin{equation}
\langle\delta\phi(k,t)\delta\phi(k',t)\rangle\approx
                 {T\over\omega_k^2}\,(2\pi)^2\delta^{(3)}(k+k'),\label{ss1}
\end{equation}
where $\omega_k^2=k^2+m^2$. This is the correct statistical mechanical
result, telling us that the oscillator modes with energy
$\omega^2_k\,\delta\phi^2$ have an average energy $T$ in the classical
regime $\omega_k \ll T$.  In the quantum regime $\omega_k \gg T$, we
would have

\begin{equation}
\langle\delta\phi(k,t)\delta\phi(k',t)\rangle\approx {1\over
  2\omega_k}\,(2\pi)^2\delta^{(3)}(k+k').\label{ss2}
\end{equation}
This result can be obtained by following the general prescription
(see, e.g., \cite{Ramos:2013nsa} where this is explicitly
derived) of inserting the factor $(\omega/2T)\cosh(\omega/2T)$ into
the Fourier transform of the noise correlation (\ref{phin}).


\section{cosmological perturbations}
\label{sec3}

In this section we shall describe the effects of fluid and scalar
field fluctuations in a cosmological setting where the background
spacetime describes a homogeneous, isotropic and spatially flat
universe.  We assume the fluid to be highly relativistic, such as we
might expect in the very early universe.  The main dissipative
mechanisms are the energy loss by the scalar field and viscosity in
the radiation fluid. Each of these is associated with a stochastic
source term with correlation functions determined by the
fluctuation-dissipation relation. We shall take the damping terms and
the correlation functions to have a local form, allowing us to apply
the equivalence principle.

Our gauge-ready formalism for cosmological perturbations follows Hwang
and Noh \cite{hwangnoh}.  The spacetime metric for a scalar-type of
perturbation is given by

\begin{equation}
ds^2=-(1+2\alpha)dt^2-2\beta_{,i}dt \,
dx^i+a^2\left(\delta_{ij}(1+2\varphi)+2\gamma_{,ij}\right)
dx^idx^j,\label{metric}
\end{equation}
where $a(t)$ is the scale factor and $H=\dot a/a$ defines the
background expansions rate.  Physical combinations of the metric
perturbations which will be useful later on are the shear $\chi$ and
perturbed expansion rate $\kappa$, given by

\begin{eqnarray}
\chi&=&a(\beta+a\dot\gamma),\\ \kappa&=&3H\alpha-3\dot\varphi-\nabla^2\chi.\label{defkappa}
\end{eqnarray}
The background Laplacian denotes the combination
$\nabla^2=a^{-2}\delta^{ij}\nabla_i\nabla_j$, where $\nabla_i$ is the
derivative with respect to $x^i$. Note the factor of $a^{-2}$ here,
and that $\nabla^2$ is the covariant Laplacian for the spatial metric
$g_{ij}=a^2\delta_{ij}$

The stress-energy tensor is conveniently expressed as,

\begin{equation}
T_{ab}=(\rho+p)n_an_a+p g_{ab}+n_aq_b+n_bq_a+\Pi_{ab},
\label{set1}
\end{equation}
where $q_a$ and the trace-free tensor $\Pi_{ab}$ are orthogonal to the
unit vector $n_a$.  We shall take $n^a$ to be the unit normal to the
constant-time surfaces.  {}For scalar perturbations, we replace $\rho$
by $\rho+\delta\rho$, $p$ by $p+\delta p$ and define $\delta v$ and
$\delta\Pi$ by

\begin{equation}
q_i=(\rho+p)\nabla_i \delta v,\qquad
\delta\Pi_{ij}=\nabla_i\nabla_j\delta\Pi-\frac13g_{ij}\nabla^2\delta\Pi,
\label{defpi}
\end{equation}
The perturbed Einstein equations in gauge-ready form are
then~\cite{hwangnoh}

\begin{eqnarray}
\nabla^2\varphi+H\kappa&=&-4\pi
G\delta\rho,\label{e1}\\ \kappa+\nabla^2\chi&=&-12\pi G(\rho+p)\delta
v,\label{e2}\\ \dot\chi+H\chi-\alpha-\varphi&=&8\pi
G\delta\Pi,\label{e3}\\ \dot\kappa+2H\kappa+\nabla^2\alpha-3(\rho+p)\alpha&=&4\pi
G(\delta\rho+3\delta p).\label{e4}
\end{eqnarray}
Diffeomorphism invariance allows us to fix two of the independent
variables.  At least two further equations are required, and these
come from considering the matter sector, which in our case consists of
the radiation fluid and the scalar field.

\subsection{Fluid and scalar perturbations}

The stress-energy tensor can be expressed in the velocity frame we
used earlier in Eq. (\ref{set}),

\begin{equation}
T_{ab}=Ts\,u^{(f)}_a u^{(f)}_b-\Omega \,g_{ab}+\Pi^{(f)}_{ab}
+\nabla_a\phi\nabla_b\phi-\frac12(\nabla\phi)^2g_{ab}.\label{set2}
\end{equation}
We have taken the energy flux $\boldsymbol{P}=0$ to simplify the
discussion, but non-vanishing energy flux can easily be accommodated.
By comparing the two forms of the stress-energy tensor (\ref{set1})
and (\ref{set2}) on homogeneous backgrounds with vanishing fluid
velocity we find the relations,

\begin{eqnarray}
\rho&=&Ts+\frac12\dot\phi^2+\Omega,\\ p&=&\frac12\dot\phi^2-\Omega+\Pi^{(f)},
\end{eqnarray}
where $\Pi^{(f)}=\Pi^{(f)}{}_i{}^i$.  {}For the fluctuations,
comparing the first-order perturbations of the two stress-energy
tensors gives

\begin{eqnarray}
\delta\rho&=&\delta\rho^{(f)}+\dot\phi(\delta\dot\phi-\alpha\dot\phi)+\Omega_{,\phi}\delta\phi,
\label{drp}\\
\delta p&=&\delta
p^{(f)}+\dot\phi(\delta\dot\phi-\alpha\dot\phi)-\Omega_{,\phi}\delta\phi,\\ (\rho+p)\delta
v&=&Ts\,\delta v^{(f)}-\dot\phi\,\delta\phi,
\label{drv}
\end{eqnarray}
where $\delta\rho^{(f)}=T\delta s$ and $\delta
u^{(f)}_i=\nabla_i\delta v^{(f)}$ as before. (See below for $\delta
p^{(f)}$.)

\subsection{Fluid equations}

The fluid equations obtained from the stress-energy tensor
(\ref{set2}) are

\begin{eqnarray}
T\,Ds+Ts\,\nabla_a
u^{a(f)}+\Pi^{(f)ab}\nabla_au_b{}^{(f)}&=&-Q^{(\phi)},
\label{fullfluid1}\\
Ts\,Du_a{}^{(f)}+s\,h_a{}^b\nabla_bT+h_{ac}\nabla_b\Pi^{(f)bc}&=&-h_{ac}Q^{(\phi)c}.
\label{fullfluid2}
\end{eqnarray}
where $h_{ab}=g_{ab}+u_a{}^{(f)}u_b{}^{(f)}$ and $D$ is the comoving
derivative as before.  Guided by the equivalence principle, we add
dissipation and noise sources to the shear stress $\Pi^{(f)}{}_{ab}$
to reproduce the flat-spacetime limit Eq. (\ref{sst0}),

\begin{equation}
\Pi^{(f)}{}_{ab}=-2\,\eta_s\,\sigma_{ab}-\eta_b
h_{ab}\nabla^{c}u_c{}^{(f)}-\Sigma_{ab}.\label{constit}
\end{equation}
The first term relates the shear stress to the rate-of-strain tensor
$\sigma_{ab}$,

\begin{equation}
\sigma_{ab}=h_{(a}{}^ch_{b)}{}^d\nabla_{c}u_d{}^{(f)}-\frac13h_{ab}\nabla^{c}u^{(f)}_{c}.
\end{equation}
Note that the bulk viscosity terms behave like a contribution to the
pressure $p^{(f)}$.

We are ready to expand these equations to first order in perturbations
theory about homogeneous backgrounds. The background fluid equation
from Eq.~(\ref{fullfluid1}) is

\begin{equation}
T\dot s+3H(Ts-3H\eta_b)=\Upsilon\dot\phi^2.
\end{equation}
At first order in perturbation theory, using the metric
(\ref{metric}), we find the velocity expansion,

\begin{equation}
\delta(\nabla^{c} u^{(f)}_{c})=\nabla^2\delta v^{(f)}-\kappa,
\end{equation}
and the strain tensor

\begin{equation}
\sigma_{ij}=\nabla_i\nabla_j\sigma-\frac13g_{ij}\nabla^2\sigma,\qquad
\sigma=\delta v^{(f)}+\chi.\label{defsig}
\end{equation}
We can also modify the pressure to absorb the bulk viscosity, by
defining

\begin{equation}
\delta p^{(f)}=s\delta
T-3H\delta\eta_b=c_s^2(\delta\rho^{(f)}-Ts_{,\phi}\delta\phi)
-3H\delta\eta_b,
\end{equation}
after taking into account Eq. (\ref{dpp}).  The fluid equations
(\ref{fullfluid1}) and (\ref{fullfluid2}) expanded to first order
with the metric (\ref{metric}) become

\begin{eqnarray}
\delta\dot\rho^{(f)}-\alpha T\dot s+3H(\delta\rho^{(f)}+\delta
p^{(f)}-\eta_b\kappa)+(Ts-3H\eta_b) (\nabla^2\delta
v^{(f)}-\kappa)+s_{,\phi}\delta q&=&-\delta Q^{(\phi)},
\label{fluid1}\\
 a^{-3}\{a^3(Ts-3H\eta_b)\delta
 v^{(f)}\}\dot{\phantom{a}}+\alpha(Ts-3H\eta_b) +\delta
 p^{(f)}-\eta_b\kappa-\eta^\prime\nabla^2(\delta
 v^{(f)}+\chi)&=&-\delta J^{(\phi)} +(2\eta^\prime T)^{1/2}\xi^{(f)}.
\label{fluid2}
\end{eqnarray}
These equations reduce to the previous set of equations (\ref{fsf1})
and (\ref{fsf2}) in flat space if we substitute
$Ts=\rho^{(f)}+p^{(f)}$, although it is often advantageous to work
with $s$ and $T$ rather than $\rho^{(f)}$ and $p^{(f)}$.

The correlation function for the stochastic sources has to be
corrected to account for the scaling between comoving coordinates
$x^i$ and inertial frame coordinates $a x^i$, resulting in a factor
$a^{-3}$,

\begin{equation}
\langle\xi^{(f)}(x,t)\xi^{(f)}(x',t')\rangle=a^{-3}\delta^{(3)}(x-x')\delta(t-t').
\label{cnoise}
\end{equation}
  The energy and momentum transfer terms are given as before by
  perturbing Eq.~(\ref{flux}),

\begin{eqnarray}
\delta Q^{(\phi)}&=&-\delta\Upsilon\dot\phi^2
-2\Upsilon\dot\phi\,(\delta\dot\phi-\alpha\dot\phi) +(2\Upsilon
T)^{1/2}\dot\phi\,\xi^{(\phi)}+\boldsymbol{\nabla}\cdot\boldsymbol{P},
\label{qf}\\
\delta J^{(\phi)}&=&\Upsilon\dot\phi\,\delta\phi
+\nabla^{-2}\boldsymbol{\nabla}\cdot(\boldsymbol{\dot
  P}+4H\boldsymbol{P}),
\label{jf}
\end{eqnarray}
where we have allowed for the possibility of modifying the stress
energy tensor by including a stochastic energy flux term
$\boldsymbol{P}$.

The fluid equations reduce to previously known versions in special
cases.  The equations agree with other work on random radiation fluids
when the scalar field is absent and $\delta Q^{(\phi)}=\delta
J^{(\phi)}=s_{,\phi}=0$ \cite{zimdahl2}.  The non-viscous case
$\eta_s=\eta_b=0$ has been widely discussed in the context
warm inflation \cite{hmb,warmpert}. A new feature of these equations
is the noise term in the energy and momentum transfer terms (\ref{qf})
and (\ref{jf}), and the effect on the amplitude of density
perturbations will be analyzed later. The viscous case without the
random fluid sources has been discussed in \cite{shear,bulkback}.

\subsection{Scalar equation}

The perturbed version of the Langevin equation of the scalar field is
constructed along similar lines. We employ the equivalence principle
to infer the curved space Langevin equation
\begin{equation}
-\Box\phi(x,t)+\Upsilon D\phi(x,t)+\Omega_{,\phi} = (2\Upsilon
T)^{1/2}\xi^{(\phi)}(x,t),\label{stoc2}
\end{equation}
After replacing the scalar field by $\phi+\delta\phi$ and using the
perturbed metric (\ref{metric}) to get the first order perturbation
equation

\begin{equation}
(\delta\dot\phi-\alpha\dot\phi)\dot{\phantom{a}}+3H(\delta\dot\phi-\alpha\dot\phi)
  -\nabla^2\delta\phi+\Omega_{,\phi\phi}\delta\phi-\kappa\dot\phi
  +\delta\Upsilon\dot\phi+\Upsilon\delta\dot\phi-\alpha\ddot\phi
  =(2\Upsilon T)^{1/2}\xi^{(\phi)}.
\label{scalar}
\end{equation}
The correlation function for the stochastic sources has to be
corrected for the shift to comoving coordinates, as we did for the
fluid,

\begin{equation}
\langle\xi^{(\phi)}(x,t)\xi^{(\phi)}(x',t')\rangle=a^{-3}\delta^{(3)}(x-x')\delta(t-t').\label{snoise}
\end{equation}
Note that we have made no changes to the noise coefficient in
Eq. (\ref{scalar}) compared to the one used in flat spacetime,
assuming that the noise correlation functions are local and subject to
the equivalence principle. However, the effective damping term in the
equation is no longer $\Upsilon$ but $\Upsilon+3H$. Consequently, the
scalar field correlations are smaller than they would be in flat
spacetime. We interpret this effect as being due to the scalar field
correlations being non-local in time and sensitive to the finite
timescale $H^{-1}$ set by the expansion. As a result, if $\Upsilon \ll
H$, it is possible for the thermal fluctuations in the scalar field to
become smaller than the quantum fluctuations, which are at least of
order $H^2$.

The noise terms take no account of the quantum vacuum fluctuations and
the thermal fluctuations \cite{Berera:1995wh} of the inflaton, and
therefore the formalism has to be modified to properly describe the
$\Upsilon \ll H$ limit.  One way to do this through the noise terms
has been described in \cite{Ramos:2013nsa}, and we will use this
in Sect \ref{sec5}. Another approach is to make use of the linearity
of the perturbation equations to add the vacuum fluctuations in as an
initial condition. Since the quantum modes and the fluctuations both
evolve by the same homogeneous equations, this will reproduce the
quantum vacuum contribution correctly.

\subsection{Gauge-invariant variables}

After choosing a gauge, Eqs. (\ref{fluid1}), (\ref{fluid2}) and
(\ref{scalar}) can be combined with any two of the Einstein equations
(\ref{defkappa}) or (\ref{e1}-\ref{e3}) to form a complete
system. {}For example, in constant-curvature gauge $\varphi=0$,
denoted by a subscript $\varphi$, Eqs. (\ref{defkappa}), (\ref{e1})
and (\ref{e2}) imply

\begin{eqnarray}
H\alpha_\varphi&=&-4\pi G (\rho+p)\delta
v_\varphi,\label{ccg1}\\ H\kappa_\varphi&=&-4\pi
G\delta\rho_\varphi,\label{ccg2}
\end{eqnarray}
These can be used together with Eqs. (\ref{fluid1}), (\ref{fluid2})
and (\ref{scalar}) for the fluid and scalar
perturbations. Alternatively, in constant-shear gauge $\chi=0$, denoted
by a subscript $\chi$, Eqs. (\ref{e2}) and (\ref{e3}) imply

\begin{eqnarray}
\kappa_\chi&=&-12\pi G(\rho+p)\delta
v_\chi,\\ \alpha_\chi&=&-\varphi_\chi-8\pi G\delta\Pi.
\end{eqnarray}
The noise term appears in constant-shear gauge through the
(gauge-invariant) total stress term $\delta\Pi=\delta\Pi^{(f)}$.
Comparison of Eq (\ref{constit}) with the definitions (\ref{defpi})
and (\ref{defsig}) gives

\begin{equation}
\delta\Pi=-\eta_s\,\delta v_\chi-\frac34(2\eta^\prime
T)^{1/2}\nabla^{-2}\xi^{(f)},
\end{equation}
where $\nabla^{-2}$ is the inverse Laplacian. In constant-shear gauge
we can use Eqs. (\ref{e1}), (\ref{e2}), (\ref{fluid1}) and
(\ref{scalar}) to solve for the curvature $\varphi_\chi$, density and
scalar perturbations.

Whatever gauge we choose for solving the equations, the density
perturbations can be expressed in terms of commonly used
gauge-invariant combinations.  Two popular choices are the Lukash
variable $\Phi$ and the Bardeen variable $\Psi$ \cite{kodama},

\begin{equation}
\Phi=\varphi+H\delta v, \qquad\Psi=\varphi+{\delta\rho\over
  3(\rho+p)}.\label{defgiv}
\end{equation}
We can regard these as the curvature fluctuation $\varphi_v$ in
comoving gauge $\delta v=0$ and the curvature fluctuation
$\varphi_{\delta\rho}$ in constant density gauge $\delta \rho=0$,
respectively.  The large-scale behavior of the Lukash and Bardeen
variables will play an important role in the following sections, and
so we will review the large-scale behavior next.

Using the comoving wave number $k$, we introduce a parameter
$z=k/(aH)$ which is small in the large-scale limit. The Fourier
components of $\nabla^2\varphi$ and $\nabla^2\chi$ in comoving gauge
are assumed to be of order $z^2$.  The Einstein equations (\ref{e2})
and (\ref{e1}) imply that $\kappa_v={\cal O}(z^2)$ and
$\delta\rho_v={\cal O}(z^2)$. Gauge invariance allows us to rewrite
Eqs. (\ref{defgiv}) in comoving gauge, so that on large scales

\begin{equation}
\Psi=\varphi_v+{\delta\rho_v\over 3(\rho+p)}\approx\Phi.
\end{equation}
{}Furthermore, the definition (\ref{defkappa}) and the Einstein
equation (\ref{e4}) imply

\begin{equation}
\dot\Phi=\dot\varphi_v\approx H\alpha_v\approx{3H\,\delta p_v\over
  \rho+p}\approx{3H\,e\over \rho+p},
\end{equation}
where $e=\delta p-\bar c_s^2\delta\rho$ is the gauge invariant entropy
perturbation and $\bar c_s^2=\dot p/\dot\rho$. In the absence of
entropy perturbations, we recover the well-known result that the
Lukash and Bardeen variables approach a common constant value. The
random fluid can affect the large scale behavior through the
generation of entropy fluctuations. The noise term in the fluid
equations is suppressed by the scale factor in the correlation
function (\ref{cnoise}). This is very convenient, because the noise
and damping terms depend on physical processes which cannot apply on
length scales larger than the horizon size.  Note, however, that
whilst the scalar field will always generate entropy as it decays, it
does not necessarily generate entropy fluctuations. An example of this
is warm inflation. In homogeneous warm inflationary models, the total
density and total pressure are determined by a slow roll approximation
in terms of the value of the scalar inflaton field.  On large scales,
when spatial derivatives are dropped, this is still valid and the
pressure can therefore be expressed as a function of the density,
consequently $\bar c_s^2=\partial p/\partial\rho$ and $e=0$.

\section{Dissipation and Viscosity coefficients}
\label{diss+visc}

In this section we give generic expressions for the dissipation and
viscosity coefficients, which can be derived once a specific model for
the scalar field and its coupling to other fields is specified.

\subsection{The Langevin equation of motion for the
scalar field and the dissipation coefficient}

The derivation of the Langevin equation of motion for the scalar field
eq.~(\ref{stoc}) typically starts by setting all fields in the context
of the so called in-in, or Schwinger closed-time path functional
formalism~\cite{Hu-book}. In this formalism time dependence of the
quantities and nonequilibrium evolution can be properly described. In
particular, the dissipation and the stochastic noise source terms
appearing in eq.~(\ref{stoc}) can be both determined.  In the
closed-time path formalism the time integration is along a contour in
the complex time plane, going from $t=- \infty$ to $+ \infty$ (forward
branch) and then back to $t=-\infty$ (backwards branch). Fields are
then identified on each of the time branches like, e.g., $\phi_1$ and
$\phi_2$, respectively, and so on for all other fields in the
system. Due to the duplication of field variables in this formalism,
four two-point Green functions can be constructed with each of these
fields. In addition, it is also convenient in this formalism to work
in a rotated basis for the field variables, called the Keldshy basis,
where we define new field variables: $\phi_{c} = (\phi_1 + \phi_2)/2$
and $\phi_{\Delta} = \phi_1 - \phi_2$. The effective equation of
motion for the scalar field is obtained from the saddle point
equation~\cite{GR}

\begin{equation}
\frac{\delta \Gamma[\phi_c,\phi_\Delta] }{\delta \phi_{\Delta} }
\Bigr|_{\phi_\Delta=0} = 0\;,
\label{eom}
\end{equation}
where $\Gamma[\phi_c,\phi_\Delta]$ is the effective action for the
scalar field. This leads to an effective Langevin-like stochastic
equation of motion of the form~\cite{GR,Berera:2008ar}
\be
\int d^4 x' {\cal O}_R[\phi_{c}](x,x') \phi_{c}(x') = \xi(x) \;,
\label{langevin}
\ee
where ${\cal O}_R[\phi_{c}](x,x')$ is defined as

\begin{equation}
{\cal O}_R[\phi_{c}](x,x') = \left[\partial^2 + V''(\phi_c) +
  \Sigma_{R,{\rm local}}\right] \delta^4(x-x') +
\Sigma_{R}[\phi_c](x,x')\;,
\label{ORchi}
\end{equation}
where $\Sigma_{R}[\phi_c](x,x')$ are retarded corrections coming from
the functional integration that leads to the effective action in the
Keldshy basis and $\Sigma_{R,{\rm local}}$ indicates local
corrections. The term on the right hand side in eq.~(\ref{langevin}),
can be interpreted as a Gaussian stochastic noise with the general
properties of having zero mean, $\langle \xi(x) \rangle =0$, and
two-point correlation

\begin{equation}
\langle \xi(x) \xi(x') \rangle = \Sigma_F[\phi_c] (x,x')\;,
\label{noisetwopoint}
\end{equation}
where $\Sigma_F[\phi_c] (x,x')$ is the diagonal self-energy in the
Keldshy basis.

Equation~(\ref{langevin}) is a nonlocal, non-Markovian equation of
motion for $\phi$.  It can be shown though that when there is a clear
separation of timescales in the system, this equation can be well
approximated by a local, Markovian approximation~\cite{BMR}. In this
case, the eqs.~(\ref{langevin}) and (\ref{noisetwopoint}) becomes of
the form of eqs.~(\ref{stoc}) and (\ref{phin}), with a local
dissipation coefficient that is defined
by~\cite{mossxiong,Berera:2008ar}

\begin{equation}
\Upsilon= \int d^4 x' \, \Sigma_R[\phi_c](x,x') \, (t'-t) \;.
\label{Upsilon}
\end{equation}
{}For example, if we consider an interaction term of the scalar field
$\phi$ with some other scalar field $\chi$ as given by $(g^2/4) \phi^2
\chi^2$, then the $\Sigma_R[\phi_c](x,x')$ term in eq.~(\ref{Upsilon})
becomes

\begin{equation}
\Sigma_R[\phi_c](x,x') = -i g^4 \phi_c^2 \; \theta(t-t')
\langle[\chi^2(x),\chi^2(x')]\rangle\;.
\end{equation}
Complete expressions for $\Upsilon$ can be found, e.g., in
ref.~\cite{BasteroGil:2010pb,BasteroGil:2012cm} for different
interactions and regimes of parameters.

\subsection{The viscosity coefficients}
\label{visccoefs}

The shear and bulk viscosity coefficients have been computed in
previous works and defined through Kubo formulas~\cite{kubo}, which
are derived in the context of linear response theory (see
also~\cite{zubarev}):

\begin{eqnarray}
\eta_s &=& \frac{1}{20} \lim_{\omega\to 0}\frac{1}{\omega} \int d^3x
dt e^{ i \omega t} \langle [\Pi_{lm}({\bf x},t) , \Pi^{lm}(0)]
\rangle\;,
\label{shearlr}
\\ \eta_b &=& \frac{1}{2} \lim_{\omega\to 0}\frac{1}{\omega} \int d^3x
dt e^{ i \omega t} \langle [{\cal P} ({\bf x},t), {\cal P} (0)]
\rangle\;,
\label{bulklr}
\end{eqnarray}
where

\begin{eqnarray}
\Pi_{lm}(x) &=& T_{lm}(x) - \frac{1}{3} \delta_{lm} T_i^{\;i}(x)\;,
\label{Pilm}
\end{eqnarray}
is the traceless part of the stress tensor and

\begin{eqnarray}
{\cal P}(x) &=& -\frac{1}{3} T_i^{\;i}(x) + v_s^2 T_{00}(x)\;,
\label{calP}
\end{eqnarray}
where $v_s$ is the local (equilibrium) speed of sound (introduced
explicitly in the quantum field theory calculation for consistency,
see e.g. \cite{jeon,arnold})

\begin{equation}
v_s^2 = \frac{\partial p }{\partial \rho }\;.
\label{vs}
\end{equation}
The averages in Eqs.~(\ref{shearlr}) and (\ref{bulklr}) are again with
respect to thermal equilibrium.

Typical expressions for the bulk and shear coefficients follow from
the standard hydrodynamics which expresses these coefficients in terms
of the collision time $\tau$ of the radiation bath and the radiation
energy density $\rho_r$ \cite{weinberg}

\begin{eqnarray}
&& \eta_s = \frac{4}{15} \rho_r \tau ,
\label{weinberg-shear} \\
&& \eta_b = 4 \rho_r \tau \left(\frac{1}{3} - v_s^2 \right)^2.
\label{weinberg-bulk}
\end{eqnarray}
Note that in conformal field theories $v_s^2=1/3$ and the bulk
viscosity vanishes identically.  This is because dilatation is a
symmetry and the fluid remains always in equilibrium. Likewise, for
scale invariant field theories, for an ideal equation of state,
$\omega_r=1/3$, the bulk viscosity also vanishes. But quantum
corrections in quantum field theories in general break scale
invariance (where the renormalization group $\beta$-function is
nonvanishing) and the bulk viscosity is nonvanishing as well.  The
bulk viscosity becomes directly proportional to the measure of
breaking of scale-invariance.  This is the case of standard scalar and
gauge field theories in general and these are the type of field
theories we consider to describe the particles in the radiation bath
in a microscopic context.  In any case, the bulk viscosity is expected
to be smaller than the shear.

It is useful to give the viscosity coefficients for an explicitly
example of quantum field theory, e.g., for a self-interacting quartic
scalar field model, $\lambda_\sigma \sigma^4/4 !$, where these
viscosities where derived in \cite{jeon}.  {}From the results
obtained in \cite{jeon}, the bulk viscosity for the case of the scalar
quartic self-interaction in the weak interaction regime is given by

\be
\eta_b \simeq  \left\{
\begin{array}{ll}
& 5.5 \times 10^4 \frac{\tilde{m}_\sigma^4
    m_\sigma^2(T)}{\lambda_\sigma^4 T^3} {\rm ln}^2\left[ 1.2465
    m_\sigma(T)/T\right] , \;\; m_\sigma \ll T \ll
  m_\sigma/\lambda_\sigma\\ & 8.9 \times 10^{-5} \lambda_\sigma T^3
  {\rm ln}^2 (0.064736 \lambda_\sigma), \;\; T \gg
  m_\sigma/\lambda_\sigma ,
\end{array}
\right.
\label{bulksigma}
\ee
while the shear viscosity is the same in the two temperature regimes
given in Eq. (\ref{bulksigma}),

\begin{equation}
\eta_s \simeq 3.04 \times 10^{3} \frac{T^3}{\lambda_\sigma^2}\;,
\label{shearsigma}
\end{equation}
where, in the above expressions, $m_\sigma(T)$ is the $\sigma$ scalar
field thermal mass, $m_\sigma^2(T) = m_\sigma^2 + \lambda_\sigma
T^2/24\left[ 1+ {\cal O}(m_\sigma/T) \right]$, $\tilde{m}_\sigma^2
=m_\sigma^2(T)-T^2(\partial m_\sigma^2(T)/\partial T^2) \simeq
m_\sigma^2 - \beta(\lambda_\sigma) T^2/48$, where
$\beta(\lambda_\sigma) = 3 \lambda_\sigma^2/(16 \pi^2)$ is the
renormalization group $\beta$-function.

\subsection{Perturbations of the Dissipative and Viscosity Coefficients}

To complete the specification of the fluctuation equations, we need
$\delta \Upsilon$ and $\delta \eta_b$, the fluctuations of the
dissipation and bulk viscosity coefficient.  {}For a general
temperature $T$ and field $\phi$ dependent dissipative coefficient,
given by

\begin{equation}
\Upsilon = C_\phi \frac{T^c}{\phi^{c-1}} \,,
\label{UpsilonT}
\end{equation}
we obtain that

\begin{equation}
\delta \Upsilon = \Upsilon \left[c \frac{\delta T}{T} - (c-1)
  \frac{\delta \phi}{\phi} \right] \,. \label{dupsilon}
\end{equation}
Likewise, the quantum field derivations for the bulk and shear
viscosity coefficients, $\eta_b$ and $\eta_s$, respectively, show that
they can be parametrized in the form

\begin{eqnarray}
&&\eta_b = C_b T^{d}/m_r^{d-3}\,,
\label{zetab}\\
&&\eta_s = C_s T^{s}/m_r^{s-3}\,,
\label{zetas}
\end{eqnarray}
where $m_r$ is just a constant mass scale (typically the renormalized
bare mass for the particles in the radiation bath, for example, from
the expressions in Subsec. \ref{visccoefs}, $m_r \equiv m_\sigma$).
The temperature exponents $d$ and $s$ for the bulk and the shear
viscosity coefficients are given by the specific quantum field theory
model realization describing the particles in the thermal bath and the
specific parameter regime in study. {}For example, from the
expressions (\ref{bulksigma}) and (\ref{shearsigma}) for the viscosity
coefficients derived from a thermal $\lambda_\sigma \sigma^4$ scalar
field model, which is the relevant case for warm inflation model
building, we have $d=s=3$ in the high temperature regime $T \gg
m_\sigma/\lambda_\sigma$. In this work we will be working with this
temperature dependence for both the bulk and shear viscosity
coefficients.

As far the perturbations are concerned, we should also note that a
bulk viscous pressure is a background quantity, while a shear viscous
pressure is a perturbation quantity originating from momentum
perturbations.  Thus, we only need to account for perturbations of the
bulk viscous pressure.  {}From Eq. (\ref{zetab}), the perturbation of
the bulk viscosity, $\delta \eta_b$, we have, similarly as for the
dissipation coefficient, that

\begin{equation}
\frac{\delta \eta_b}{\eta_b}= d \frac{\delta T}{T} \,.
\label{dzetab} 
\end{equation} 

Although dissipation implies departures from thermal equilibrium in
the radiation fluid, the system has to be close-to-equilibrium for the
calculation of the dissipative coefficient to hold, therefore we
assume $p_r \simeq \rho_r/3$, {\it i.e.} we consider $\omega_r=1/3$.
Using $\omega_r=1/3$, then we have that $\rho_r \propto T^4$.  Thus,
$\delta T$ appearing in Eqs. (\ref{dupsilon}) and (\ref{dzetab}), can
be expressed in terms of the radiation energy density and its
perturbation as

\begin{equation}
\frac{\delta T}{T} \simeq \frac{1}{4} \frac{\delta \rho_r}{ \rho_r}\,.
\label{deltaT}
\end{equation}


\section{Cosmological perturbations during inflation}
\label{sec5}

In warm inflation ~\cite{Berera:1995ie,wi2} (for earlier related work,
see for instance \cite{prewarm,Moss:1985wn}) there is a non-negligible
contribution from the radiation bath to the power spectrum. The
radiation bath originates from the decay of fields coupled to the
inflaton field and triggered by its dynamics during inflation. Thus,
during warm inflation there is a two-component fluid made of a mixture
of the scalar inflaton field interacting with the radiation fluid, due
to the produced decay particles. Thus, density fluctuations are sourced
primarily by thermal fluctuations of the inflaton field when coupled
with the thermal radiation bath.  These are modeled by the stochastic
Langevin equation Eq. (\ref{scalar}) for the
inflaton $\phi$, with dissipative and stochastic noise terms satisfying a
fluctuation and dissipation relation. 
In fact, this stochastic equation for the inflaton field can
be completely derived from first principles, as shown e.g. in
\cite{GR,BR,BMR}.  It has also been shown in
\cite{mossgraham,MX} that it appropriately governs the evolution
of the inflaton perturbations during warm inflation.

Explicit microscopic derivations of the resulting dissipation term in
the inflaton effective dynamical evolution equation show that the
resulting two-fluid system in warm inflation is coupled
~\cite{BGR,Berera:1998px,Berera:2008ar}. This happens because both
fluid components can exchange energy and momentum through the
dissipation term.  As was shown in ~\cite{mossgraham}, the temperature
dependence of the dissipation coefficient causes a coupling between
the inflaton perturbations with those of the radiation
perturbations. This effectively leads to growing modes in the power
spectrum, which can cause considerable fine-tuning of the inflaton
potential parameters for warm inflation.  These growing modes 
get worse the larger is the power in temperature in the dissipation
term \cite{mossgraham,shear}. 
Earlier work had developed the expression for the primordial spectrum
in warm inflation \cite{hmb,warmpert}, but had not accounted for the
growing mode.

There can also be other intrinsic microscopic decay processes in the
produced radiation bath, causing it to depart from equilibrium.  These
intrinsic dissipative effects in the radiation fluid will cause it
deviated from a perfect fluid during inflation. As the radiation fluid
departs from equilibrium, pressure and momentum changes are produced
by the particle excitations and this generates viscous effects. Among
these are the bulk and shear viscous pressures.  The presence of these
viscous processes during warm inflation can control the growing mode
arising from the temperature dependent dissipative coefficient, as has
been recently studied in~\cite{shear}.  

In \cite{shear} the fluctuation spectrum in warm inflation was
studied by including only the effects of shear viscosity in the
radiation fluid and it was assumed that the bulk viscosity is much
smaller than the shear viscosity. This is the case for quasi-conformal
radiation fluids. For instance, in quantum field theory calculations
in general, e.g., in perturbative quantum cromodynamics, which
corresponds to the high-temperature quark-gluon phase in the early
universe, the bulk viscosity has been estimate to be a factor
$10^{-3}$ to $10^{-8}$ smaller than the shear
viscosity~\cite{Arnold:2006fz}).  Even though bulk viscosities in most
regimes have smaller magnitudes than shear viscosities, there are
regimes of temperature and field parameters where the bulk viscosity
can be important. For instance, close to phase transitions or phase
changes in general, it has been shown that the bulk viscosity can be
much larger in magnitude than the shear
viscosity~\cite{bulkPT}. Furthermore, the bulk viscosity, been related
to pressure fluctuations, already contributes at the background level,
while the shear viscosity, been related to momentum fluctuations,
contribute only at the perturbation level.  The effect of the bulk
viscosity in warm inflation has been studied previously in
~\cite{delCampo:2007cy,delCampo:2010by}.  These papers did not treat
shear viscosity effects, only looked at the case of constant
dissipation (thus there was no coupling of the radiation bath
perturbations with those of the field), and looked at constant bulk
viscous pressure or one proportional to the radiation energy
density. Under these simplifying assumptions, it was found in
~\cite{delCampo:2007cy,delCampo:2010by} that bulk viscous effects
could induce a variation in the power spectrum amplitude in the order
of $4\%$. However, by including the full temperature dependence for
both the dissipation and bulk viscosity terms, as motivated by
microscopic quantum field derivations, it is possible that the effect
of the bulk viscous pressure on the power spectrum can be
significantly higher.  This possibility will be analyzed here, where
both bulk and shear viscous effects are included.

However, because of the random
noise terms in the radiation fluid equations, the growing mode is not
completely eliminated. In fact, we show that
the random noise caused by the viscous radiation fluid tends to
further contribute to the curvature perturbation spectrum, causing it
to increase even in the weak dissipative regime of warm inflation
(when the inflation dissipation term is smaller than the Hubble
parameter). Besides, viscous random noise terms tend to add more power
on smaller scales, rendering the primordial spectrum blue-tilted. 
Since the random fluctuation contributions that are added to the
curvature perturbation spectrum are proportional to the viscosity
coefficients, this allow us to put strong constraints on the level of
viscosity allowed in the warm inflation scenario.

\subsection{ Primordial spectrum and spectral index}

In warm inflation, the scalar field $\phi$ is an
inflaton and the energy density is dominated by a temperature
independent potential $V(\phi)$. The inflaton decay is described by
the damping coefficient $\Upsilon(\phi,T)$. The radiation fluid
$\rho_r\equiv \rho^{(f)}$ is produced, and continually replenished by
decay of the inflaton field. 
The background fields satisfy
\begin{eqnarray}
\ddot\phi+(3H+\Upsilon)\dot\phi+V_{,\phi}&=&0, \label{ddotphi}\\ 
\dot\rho_r+4H \left(\rho_r - \frac{9}{4} H \eta_b\right) &=&
\Upsilon\,\dot\phi^2 \,,\\ 
3H^2&=& 8\pi G\rho \,.\label{hubble}
\end{eqnarray}

Prolonged inflation requires the slow-roll conditions $|\epsilon_X|\ll
1$, where $\epsilon_X = - d \ln X/Hdt$, and $X$ is any of the
background field quantities. The
background equations at leading order in the slow-roll approximation
of small $\epsilon_X$ become
\begin{eqnarray}
3H(1+Q)\dot\phi&\simeq&-V_{,\phi}, \label{sl1}\\ 
4 \rho_r&\simeq&3Q\dot\phi^2+ 9 H \eta_b,\\ 
3H^2&\simeq&8\pi GV, \label{sl2}
\end{eqnarray}
where $Q=\Upsilon/(3H)$.

We are interested in deriving the effect with a variable background on
the amplitude of the spectrum and its spectral index $n_s$.
Numerically, we have integrated Eqs. (\ref{fluid1}),(\ref{fluid2}) and
(\ref{scalar}) for a set of modes for different wavenumbers, together
with the background equations (\ref{ddotphi})-(\ref{hubble}). 
For the background evolution, we consider a quartic chaotic model
with inflationary potential  $V=\lambda \phi^4/4$. 
For the dissipative parameter, we focus on a cubic dependence with the
temperature, $c=3$, $\Upsilon = C_\phi \frac{T^3}{\phi^2}$, 
and similarly for the shear and bulk viscosities, with $\eta_s \propto
T^3$ and $\eta_b \propto T^3$. 
This is the dependence obtained when dissipation is given by the decay
into light 
degrees of freedom of a scalar massive field coupled to the
inflaton, with mass $m \simeq g \phi$
\cite{mossxiong,BasteroGil:2010pb, BasteroGil:2012cm}.
Note that the dissipation and viscosity coefficients quoted here can be
used to a
good approximation for $\phi$ and radiation fluid modes
$k \stackrel{<}{\sim} T$.  As $k$ approaches up to $T$, there will be
corrections to these quantities that can be computed \cite{BMR}
and for $k > T$ these coefficients will decrease exponentially.

For the metric perturbations, we work in the zero-shear gauge, $\chi=0$.
Using the slow-roll equations (\ref{sl1})-(\ref{sl2}), the
gauge-invariant Lukash variable $\Phi$ defined in Eq. (\ref{defgiv})
is now
\begin{equation}
\Phi=-{1\over 1+Q}\zeta^\phi-{Q\over 1+Q}\zeta^v \,, 
\end{equation}
where 
\bea
\zeta^{\phi} &=& -\varphi + H \delta \phi/ \dot \phi \,, \\
\zeta^{v} &=& -\varphi - \delta v_r  \,.
\eea
At late times, when $z\to 0$, we have $\Phi=-\zeta^\phi=-\zeta^v$.
The power spectrum is given by:
\begin{equation}
\langle\zeta^i(k,t)\zeta^i(k',t)\rangle=P^i(k,t)\,(2\pi)^3\delta^{(3)}(k+k')\,,
\end{equation}
which can then be used to obtain the power spectrum of the
gauge-invariant comoving curvature perturbations $P_{\cal R}(k)= (4 \pi) k^3
P_\Phi(k)$. The quantum statistical average is the same as the stochastic average
in our formalism.  Since the gauge-invariant perturbations are
constant on large scales, $P_{\cal R}(k,t)$ will approach a constant value
$P_{\cal R}(k)$, which we identify as the primordial amplitude of density
perturbations \cite{shear}.  We normalise the system in a periodic box
of size $l$ to replace the momentum delta-function $\delta^{(3)}(0)$
by $l^3$. The variables can be re-scaled to absorb the factor $l^3$ by
defining
\begin{equation}
\bar\zeta^i=(k/2\pi l)^{3/2}\zeta^i,\qquad \bar\xi^i=(k/2\pi
l)^{3/2}\hat\xi^i.
\label{rescale}
\end{equation}
As a result, $\bar\xi^i$ is a unit normalized random variable with
variance
\begin{equation}
\langle\bar\xi^i(t)\bar\xi^i(t')\rangle=\delta(t-t').
\end{equation}
Once the stochastic equations are solved, the power spectra are given
by
\begin{equation}
k^{3} P^i(k,t)=\langle\bar\zeta^i(k,t)\bar\zeta^i(k,t)\rangle.\label{rps}
\end{equation}
There is no residual dependence on the normalisation scale $l$.

In the chaotic quartic model, background
evolution is such that the dissipative ratio $Q$ and $T/H$ increase
during inflation, and inflation ends when the slow-roll conditions are
violated \cite{BasteroGil:2009ec,Bartrum:2013oka,Bartrum:2013fia}.
Radiation is given as usual by $\rho_r= \pi^2 g_* T^4/30$, and for the
number of relativistic degrees of freedom we take $g_*=15/4$. 
For the smaller $k$ mode considered, we set horizon crossing at 50 e-folds
before the end of inflation. For this parameter value, we
have $Q_* \simeq 10^{-7}$ as the lower value consistent with the
condition $T/H \geq 1$. But even for such a low $Q_*$ value, by the
end of inflation we have $Q_{50} > 1$. We have run the simulations
with $\lambda=10^{-14}$. The amplitude of the primordial spectrum can
be normalized to the Planck value \cite{Planck}, $P_{\cal
  R}^{1/2}=4.69\times 10^{-5}$ by slightly adjusting the value of
$\lambda$, but this will have little effect on the spectral index.

\begin{figure}[t]
\begin{center}
\begin{tabular}{ccc}
\includegraphics[width=80mm]{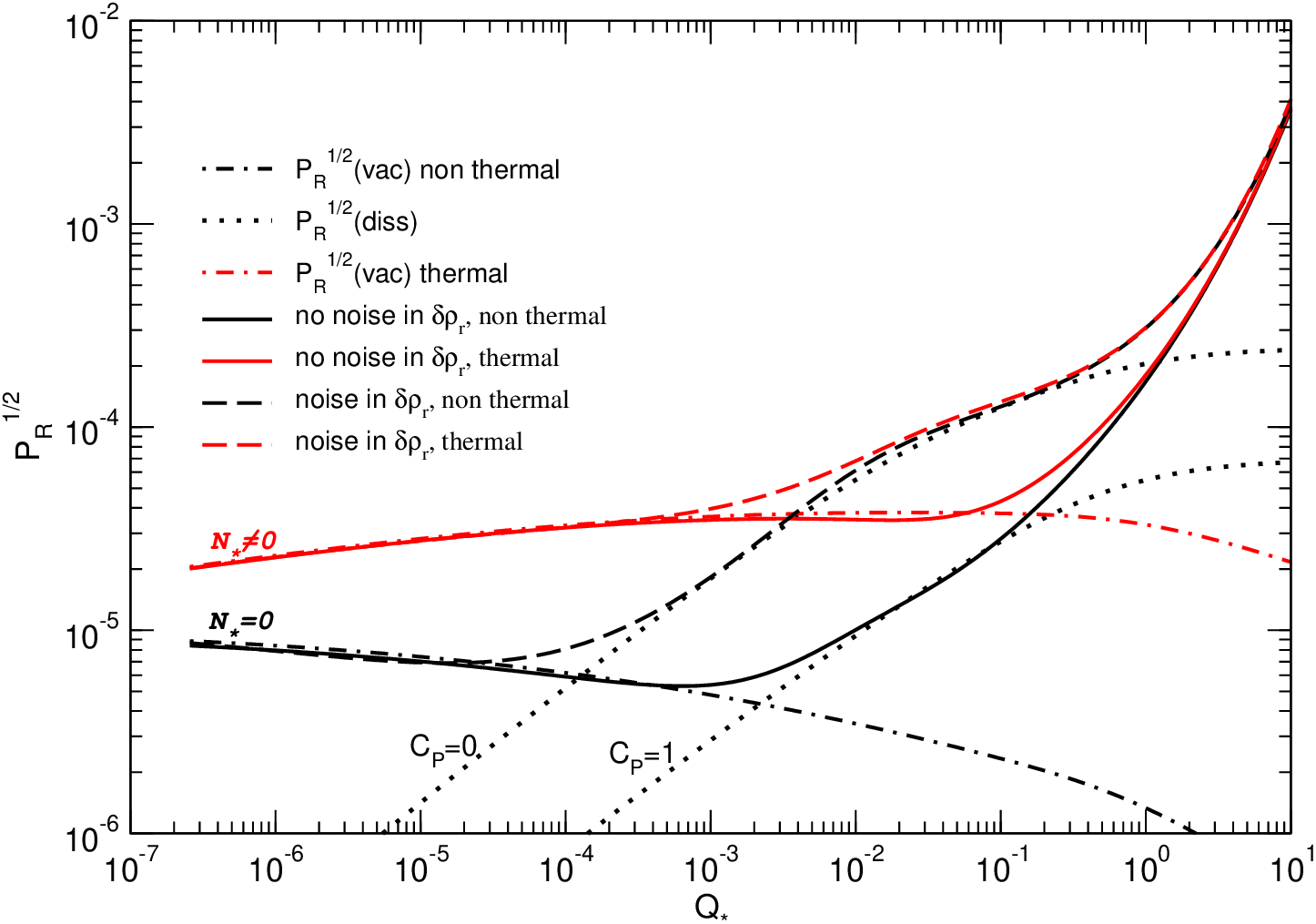} & &
\includegraphics[width=80mm]{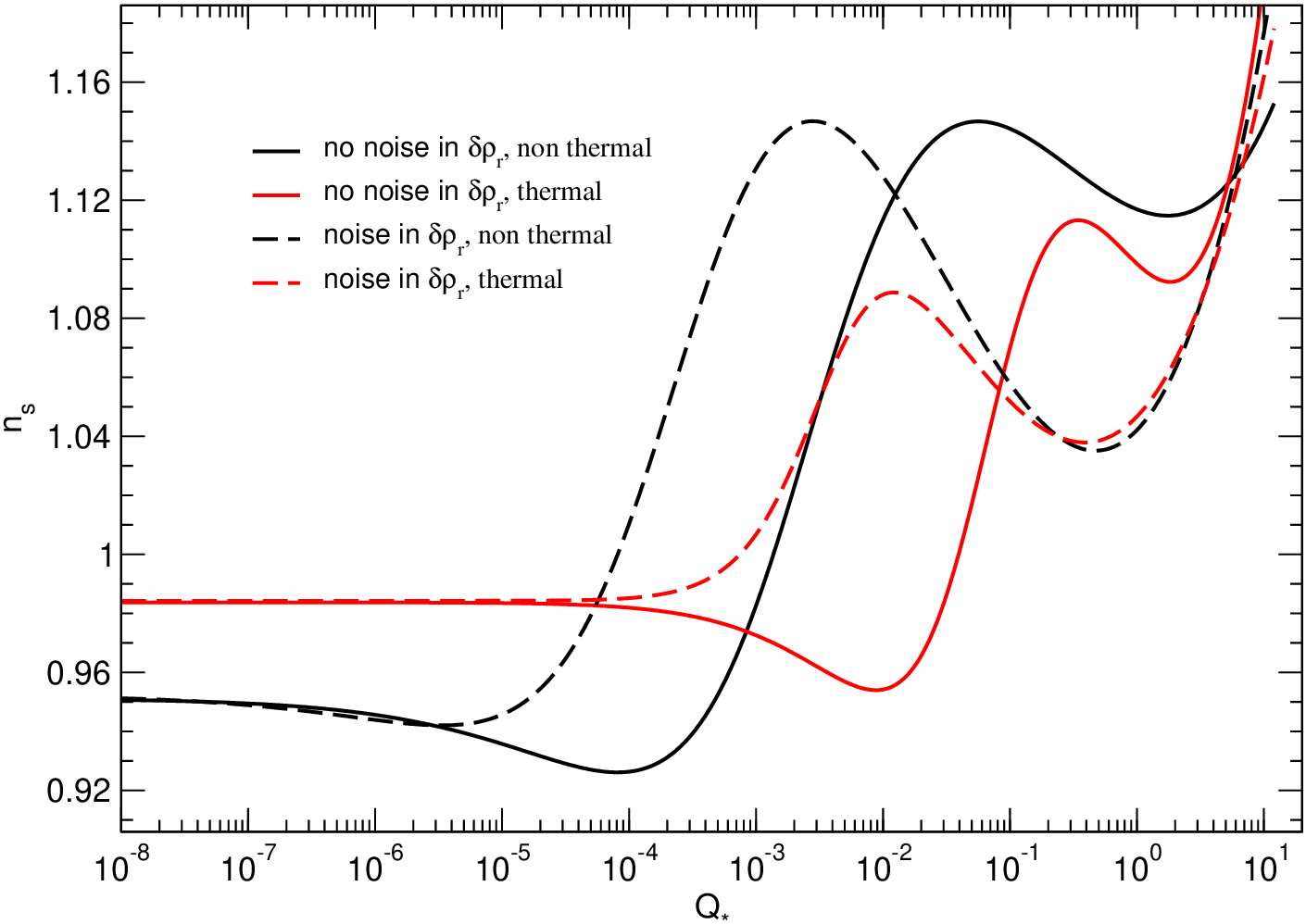}
\end{tabular}
\end{center}
\caption{LHS plot: Amplitude of the primordial spectrum versus
  $Q_*$. Solid (dashed) lines have been obtained with $C_P=1 \,(0)$;
  black lines are for ${\cal N}_*=0$ (non-thermal) and gray (red)
  lines for ${\cal N}_*=n_{BE}$ (thermal). 
  For comparison, the dotted line shows the analytical
  solution for $P_{{\cal R}, diss}^{1/2}$, and dot-dashed lines
  $P_{{\cal R}, vac}$. The amplitude of the spectrum becomes constant
  some e-folds after horizon crossing, and here we are plotting the
  value at $N_e=20$. The total number of e-folds is 50.  RHS plot:
  spectral index versus $Q_*$. We have taken $\lambda=10^{-14}$.}
\label{plotPRns} 
\end{figure}
    
Inflaton thermal and quantum fluctuations, relevant in the very weak
dissipative regime $Q_* \ll 1$, are taken into account by adding
another stochastic noise term $\xi^{(q)}$ to the field Langevin
equation, as described in \cite{Ramos:2013nsa}. Dissipative processes
may maintain a non-trivial distribution of inflation particles which
for sufficiently fast interactions should approach the Bose-Einstein
distribution $n_{BE}(k)= (e^{k/aT}-1)^{-1}$.  Both possibilities,
either negligible inflaton occupation number at horizon crossing
${\cal N}_*\simeq 0$ or given by a thermal distribution, will be
considered by adding the following stochastic term to the field
equation (\ref{scalar}):
\be 
H \frac{\sqrt{1 + 2 {\cal N}_*}}{\sqrt{2}} ~\xi^{(q)} \,, 
\ee 
with the same correlation function for the stochastic noise than that
of $\xi^{(\phi)}$ in Eq. (\ref{scalar}). 
In addition, we
have numerically explored the two possibilities encountered in section
II when discussing the mixture of a relativistic fluid and a scalar
field in a cosmological set-up: having either the field stochastic
noise $\xi^{(\phi)}$ in the energy flux ($C_P=0$) or in the momentum
flux ($C_P=1$). We focus mainly on low values of the dissipative ratio
at horizon crossing $Q_* \lesssim 10$. As we will 
see, in this regime the different interplay of the stochastic terms in
both fluid and field equations makes a difference in the spectrum and
the spectral index. For larger values of $Q_*$, the effect of the
growing mode observed in \cite{mossgraham} dominates the
spectrum.

{}For the dissipative coefficient in the weak regime, $Q_*\ll 1$, the
inflaton is effectively decoupled from the radiation fluid, and the
analytical expression for the amplitude of the primordial spectrum is
given by \cite{Ramos:2013nsa,Bartrum:2013fia}:
\be 
P_{\mathcal{R}} = ( P_{\mathcal{R},\,diss} + P_{\mathcal{R},\,vac} )
=  \left( \frac{ H_* }{\dot
  \phi_*}\right)^2 \left(\frac{H_*}{2 \pi}\right)^2 \left[
  \frac{T_*}{H_*} \frac{ 2\pi Q_*}{\sqrt{ 1 + 4 \pi Q_*/3 }} + 1+2
       {\cal N}_* \right]\,, 
\ee 
where all variables are evaluated at horizon crossing. 
The first term is the direct contribution due to
dissipation, the stochastic term $\xi^{(\phi)}$, while the the others
are due to the inflaton fluctuations including the vacuum term (${\cal
  N}_*=0$) and the contribution from thermal excitations (${\cal N}_* =
n_{BE}(a_*H_*)$). The analytical expression for
$P_{\mathcal{R},\,diss}$ was obtained for the scenario with $C_P=1$.
When the radiation source term includes the dissipative stochastic
noise, this enhances the amplitude of radiation fluctuations, which
backreacts earlier on the inflaton fluctuations, and dissipation
dominates the spectrum for smaller values of $Q_*$ (dashed lines in
the LHS plot in  Fig. (\ref{plotPRns})). Numerically, up to $Q_*
\simeq 0.1$, the contribution $P_{{\cal R},\,diss}$ gets enhanced in
this case by a factor ${\cal O}(40)$,
\be 
P_{{\cal R}\,diss}^{C_P=0} \simeq \left( \frac{ H_* }{\dot
  \phi_*}\right)^2 \left(\frac{H_*}{2 \pi}\right)^2 \frac{T_*}{H_*}
\frac{ 80 \pi Q_*}{\sqrt{ 1 + 4 \pi Q_*/3}}\,.  
\ee 
Taking this effect
into account, for small values of $Q_* \lesssim 1$ the amplitude can
be written as:
\be 
P_{\mathcal{R}} = \left( \frac{ H_* }{\dot \phi_*}\right)^2
\left(\frac{H_*}{2 \pi}\right)^2 \left[ \Delta_{Q*} + 1 + 2 {\cal N}_*
  \right]\,, \label{prweak} 
\ee 
where,
\bea \Delta_{Q*} &\simeq& 80 \pi Q_* \frac{T_*}{H_*}\,, ~~~~ C_P=0
\,,\\ \Delta_{Q*} &\simeq& 2 \pi Q_* \frac{T_*}{H_*}\,, ~~~~ C_P=1\,.
\eea 

\begin{figure}[t]
\begin{center}
\includegraphics[width=80mm]{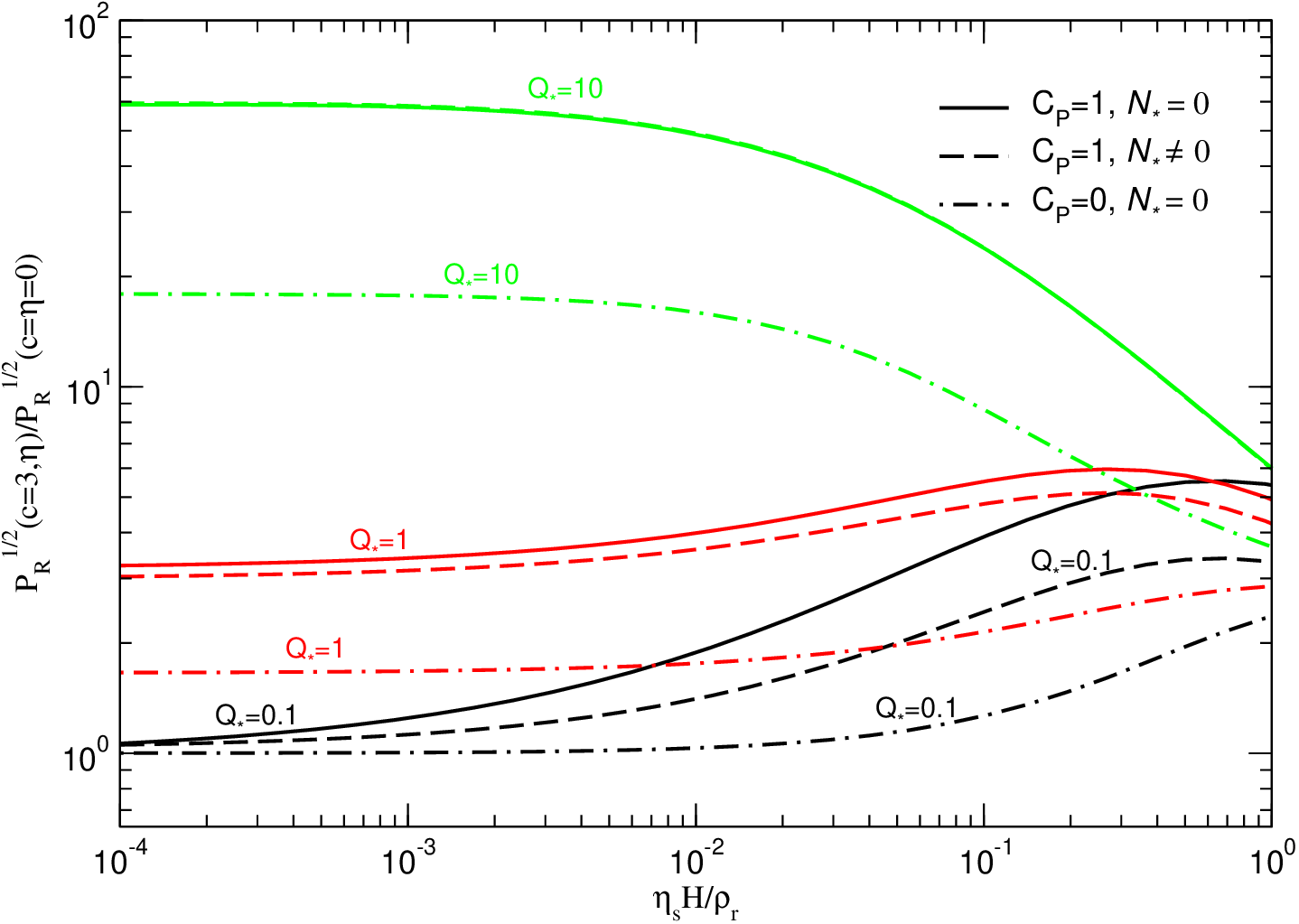} 
\end{center}
\caption{
Amplitude of the primordial spectrum versus
  $\bar \eta_{s*}=H_* \eta_{s*}/\rho_{r*}$, for different values of
$Q_*$ as indicated in the plot.  Solid (dashed) lines have been
obtained with $C_P=1$ and ${\cal N}_*=0$ (${\cal N}_* \neq0$);
dot-dashed line with $C_P=0$. In the later case, the statistical
inflaton state does not make any difference. The amplitude of the
spectrum has been normalized by its analytical value for $c=0$.   
}  
\label{plotPRshear} 
\end{figure}

The analytical expression of the spectrum matches the numerical
values up to $Q_*\lesssim 0.1$, as can be seen on the LHS plot in
Fig. (\ref{plotPRns})). For larger values, radiation back-reacts onto
the inflation fluctuations and there is a ``growing mode'' in the
spectrum, with $P_{\cal R} \propto Q_*^\alpha$.  When ${\cal
  N}_*\simeq 0$, the spectrum is dominated by the inflaton vacuum
contribution upto $Q_* \simeq 0.001 (10^{-4})$ when $C_P=1$ ($C_P=0$);
 after which dissipation takes over (dotted line) up to $Q_*\simeq 0.1$; whereas
for a thermal inflaton with ${\cal N}_* \neq 0$, the vacuum
contribution is enhanced by a factor ${\coth}(T_*/2H_*)\simeq
T_*/H_*$, and  it dominates until larger values of $Q_* \agt 0.1$. 
For values of $Q_* \agt
10$, the inflaton statistical state does not make any difference, and the
amplitude of the spectrum is fully dominated by dissipation and the
induced growing mode.  Moreover it does not make any difference
whether or not the stochastic dissipative noise sources the radiation
in the strong dissipative regime.

The spectral index is given in the companion plot in
{}Fig. (\ref{plotPRns})). Analytically, this is given by:
\be 
n_s -1 = \frac{d P_{\mathcal{R}}}{d \ln N_e} \simeq 2 \eta_* - 6
\epsilon_* +  \frac{4 {\cal N}_*}{1+2 {\cal N}_* +\Delta_{Q*}} ( 2
\epsilon_* - \eta_* + \sigma_*) + \frac{2 \Delta_{Q*}}{1+2 {\cal N}_*
  +\Delta_{Q*}} ( 7 \epsilon_* - 4 \eta_* + 5 \sigma_*) \,, 
\ee
where
\be
\epsilon = -\frac{1}{H} \frac{d \ln H}{dt} \simeq  \frac{m_P^2}{2
  (1+Q)}\left(\frac{V_{,\phi}}{V}\right)^2 ,\, ~~~~~~
\sigma= -\frac{1}{H} \frac{d \ln \phi}{dt} \simeq  \frac{m_P^2}{1+Q} \frac{V_{,\phi}/\phi}{V}
,\, ~~~~~~
\eta= -\frac{1}{H} \frac{d \ln V_{,\phi}}{dt} \simeq \frac{m_P^2}{1+Q} \frac{V_{,\phi\phi}}{V}
\ee

In particular for the quartic potential, they are given by:
\be 
\epsilon= \frac23\eta=2 \sigma = 8 \left(\frac{m_P}{\phi} \right)^2 \frac{1}{1+Q}\,,
\ee 
and the spectral index when $Q_* \ll 1$ reads: 
\be 
n_s -1 = - 3 \epsilon_* +  \frac{4 {\cal N}_*}{1+2 {\cal N}_*
  +\Delta_{Q*}}  \epsilon_*  + \frac{ 7 \Delta_{Q*}}{1+2 {\cal N}_*
  +\Delta_{Q*}} \epsilon_* \,.  
\ee

Therefore, in the very weak dissipative regime with $\Delta_{Q*}\ll
1$, we have a red-tilted spectrum: 
\bea 
n_s \simeq 1 - 3 \epsilon_* \,,~~~({\cal N}_*\simeq 0) \,,\\ 
n_s \simeq 1 -  \epsilon_* \,,~~~({\cal N}_*\simeq n_{BE})\,,  
\eea
whereas when $\Delta_{Q*} \agt 1 + 2 {\cal N}_*$ the spectrum turns
blue $n_s \simeq 1 + 4 \epsilon_*$. 
Again, this happens earlier when $C_P=0$ (dashed lines in
 Fig. (\ref{plotPRns})). Up to that point, the  spectral index is
consistent with Planck values.  When $C_P=1$ and the stochastic noise
does not source the radiation energy density fluctuations (solid
lines), the spectral index decreases before the $\Delta_{Q*}$ contribution
becomes non-negligible:  this is due to the evolution of $Q$ during
inflation in this model, always increasing. Even when small at horizon
crossing, it will become larger than one before the end of
inflation, and dissipation will dominate over the Hubble friction. We
have then 50 e-folds of inflation for smaller values of the inflaton
field, i.e., larger values of $\epsilon_*$, and thus a slightly more
red-tilted spectrum. Soon after, dissipation ($\Delta_{Q*}$) takes
over and the spectrum quickly becomes  blue-tilted. 

\begin{figure}[t]
\begin{center}
\begin{tabular}{cccc}
\includegraphics[width=80mm]{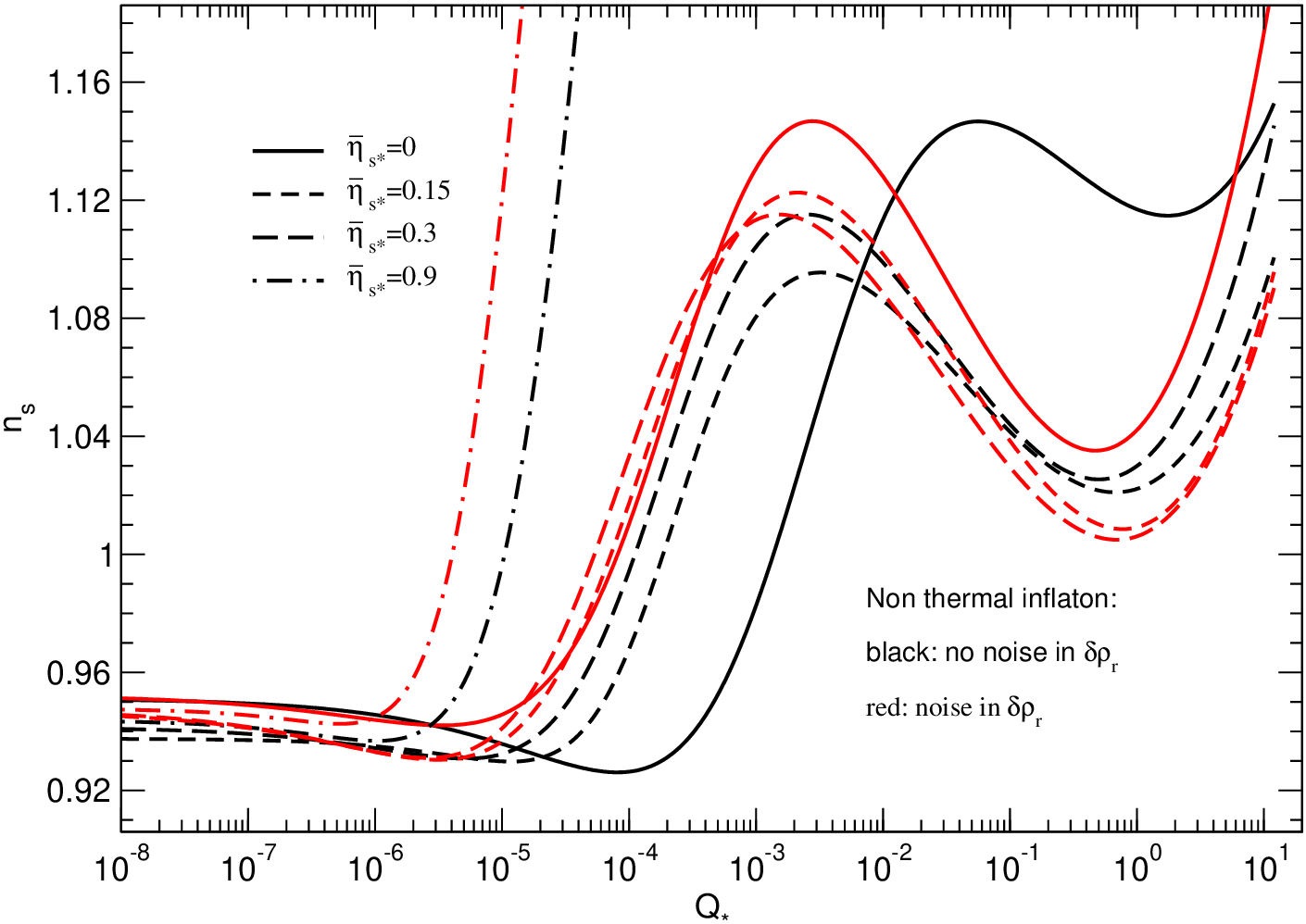} & & &
\includegraphics[width=80mm]{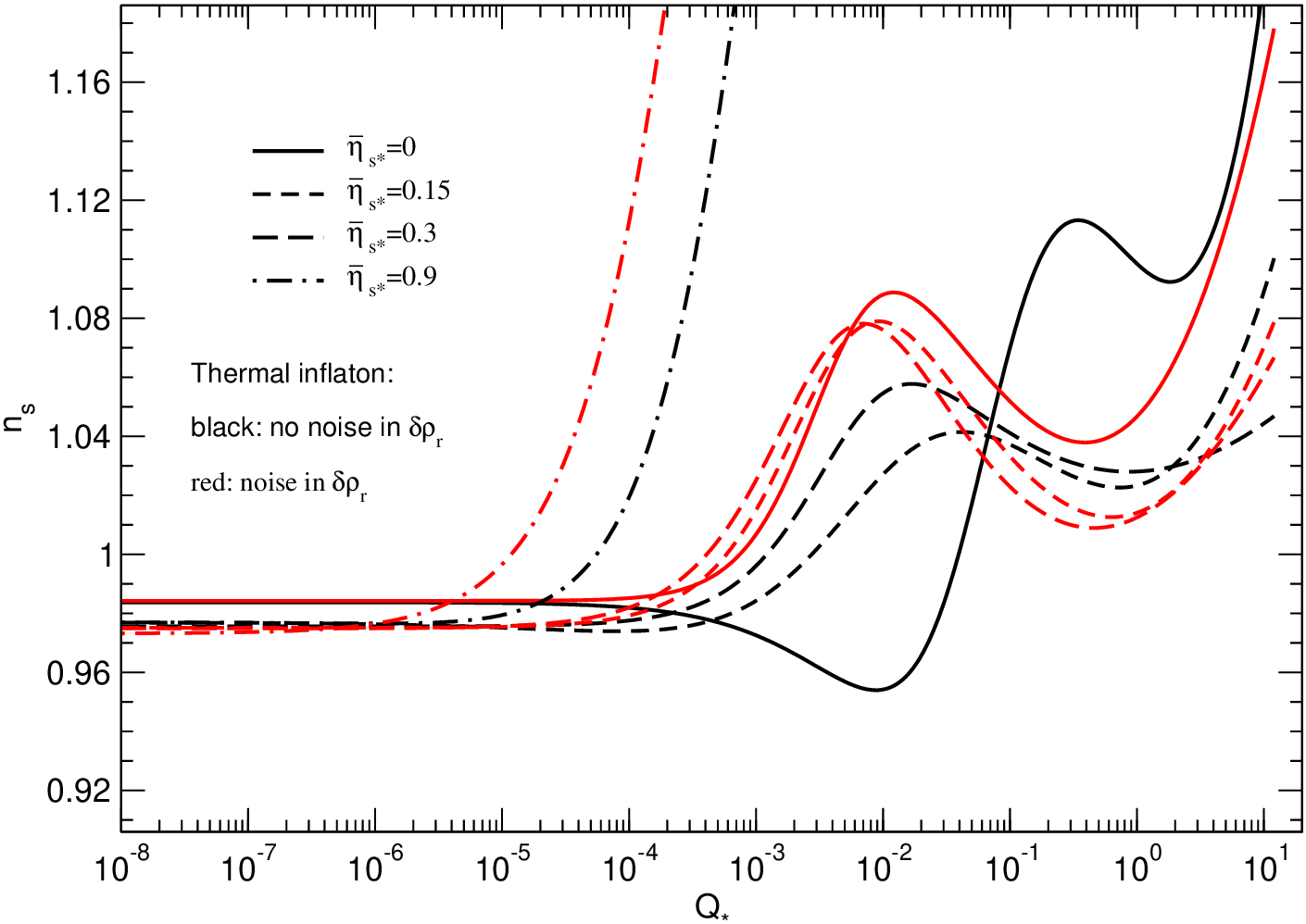}
\end{tabular}
\end{center}
\caption{RHS plot: spectral index versus $Q_*$, with a non-thermal
  inflaton, for different values of the shear parameter
  $\bar\eta_{s*}=\eta_{s*}H_*/\rho_{r*}$ at horizon crossing. Black
  lines do not include any
  noise term in the radiation fluctuations ($C_P=1$), while red lines
  do so ($C_P=0$). LHS plot: same for a thermal inflaton. We have
  taken $\lambda=10^{-14}$, and $N_*=50$.}  
\label{plotnsshear} 
\end{figure}

We now turn to the effect of the viscosities on the spectrum. 
We first set the bulk viscosity $\eta_b$ to zero, and consider the
effects of shear. In Fig. (\ref{plotPRshear}) we show the primordial
spectrum normalized by its analytical value when $c=0$
(Eq. \ref{prweak}) as a function of the shear parameter  
$\bar \eta_{s*} = H_* \eta_{s*}/(\rho_{r*})$.
Viscous effects will tend to damp down the effect of
the growing mode, although only for values $Q_* \agt {\cal
  O}(1)$; and the growing mode will only effectively disappear
for values of the shear parameter $H_* \eta_{s*}/\rho_{r*} > 1$,   
beyond the limit of validity of the assumption of being
close-to-equilibrium. For values $Q_* \leq 1$, indeed the
amplitude gets enhanced. This is due to the stochastic noise term in
the momentum fluid equation due to viscosity, which sources the fluid
momentum perturbations. This effect dominates over the friction
effect introduced by the viscosity, renders the amplitude larger, and
through the fluid energy density fluctuations will in turn affect the
field equations.

In Fig. (\ref{plotnsshear}) we have considered the effects of
the shear viscosity in the spectral index, for different values of the
shear parameter at horizon crossing.  For values $Q_* \alt 1$, shear 
will also set more power on larger
wavenumbers, implying a blue-tilted spectrum. The stochastic shear
noise effect is similar to that of the field noise when included in
the energy flux, rendering the spectrum consistent with Planck data
only in the very weak dissipative regime $Q_* \ll 1$.

\begin{figure}[t]
\begin{center}
\includegraphics[width=80mm]{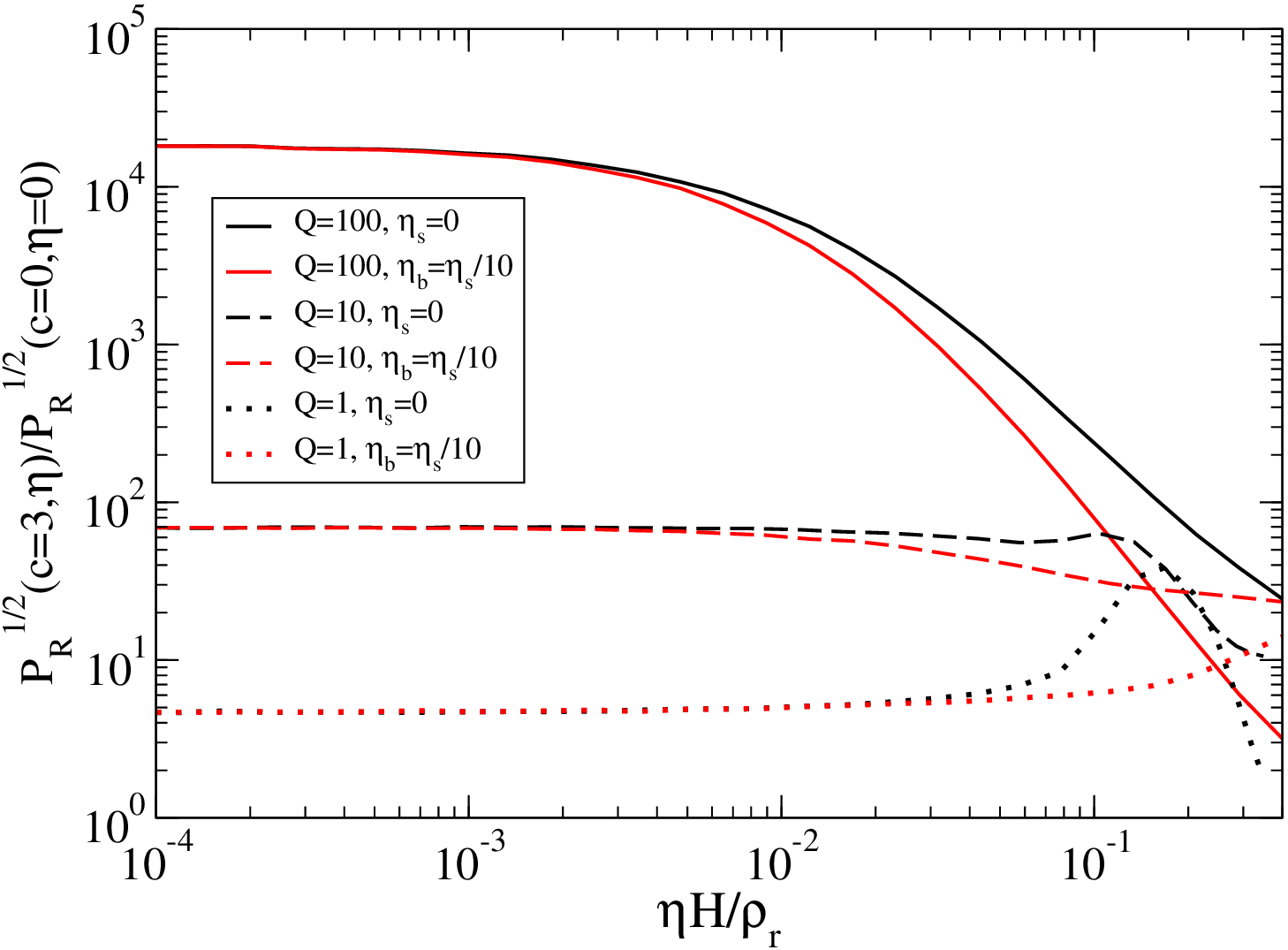}
\end{center}
\caption{The total (square root) amplitude for the power spectrum  for
  the case $c=3$ and normalized by its value for $c=0$, as a function
  of viscosity.  Three different values for the dissipation ratio $Q$
  are used to illustrate the ability of the viscosity to damp the
  spectrum. Black curves include only the effect of the bulk viscosity
  ($\eta_s=0$) and the red includes the effects of both the shear
  viscosity and bulk viscosity coefficient given by
  $\eta_b=\eta_s/10$.}
\label{plotbulk} 
\end{figure}

In  {}Fig. (\ref{plotbulk}) we show the total power spectrum as a
function of only bulk viscosity and with a combination of bulk and
shear viscosities. The results are for the case of thermalized
inflaton perturbations and for the radiation noise term ($C_P=0$).
Other cases produce results that are quantitatively not much different
than the ones shown. It is noticed that the larger is the dissipation
coefficient, the larger the amplitude of the power spectrum gets with
respect to the case where it is insensitive to magnitude of $Q$, given
by $c=0$, i.e., a temperature independent dissipation coefficient.
This is the growing mode resulting from the coupling of the inflaton
and radiation perturbations as found in ~\cite{mossgraham} and
also studied in \cite{shear}, where also the effects of shear viscosity
were considered.    {}From {}Fig.~\ref{plotbulk} we can
noticed that the combination of bulk and shear viscosities tend to
damp the spectrum quicker than including only bulk viscosity. But the
spectrum only gets effectively damped to the values where the growing
mode is compensated for values of the bulk and or shear viscosities
that are already too close to the limit of validity of the assumption
of small departures from equilibrium, i.e., $\eta H/\rho_r \ll 1$.

Results for the spectral tilt for some representative values of
dissipation coefficient and bulk viscosity are presented in
Table ~\ref{tab1}. We have included both the cases of including the
radiation noise term ($C_P=0$) and in the absence of it ($C_P=1$) in
the perturbation equation. We have included also the cases of thermal
and nonthermal inflaton fluctuations for comparison.  We find that in
general, for not too small dissipation coefficient, $Q_* \lesssim
10^{-3}$, for thermalized inflaton fluctuations and by including a
small  viscosity coefficient  $\eta_{b*} H_*/\rho_{r*} \lesssim
0.035$, the results can be rendered compatible with Planck
data.  

\begin{table}[htb]
\begin{center}
\begin{tabular}{c|c|c|c|c|c}
\hline  $Q_*$ &  \multicolumn{3}{c|}{$10^{-5}$} &
\multicolumn{2}{c}{$10^{-3}$} \\ \hline ${\bar \eta}_{b*}$ &
$2.2\times 10^{-5}$ & $0.1$ &  $0.188$ & $0.035$ & $0.21$ \\ \hline
(a) & $0.97\pm 0.02$ & $0.980 \pm 0.001$ & $1.008 \pm 0.003$ & $0.969
\pm 0.004$ & $1.89 \pm 0.01$\\ (b) & $0.98 \pm 0.01$ & $0.989 \pm
0.008$ & $1.51 \pm 0.01$ & $1.03 \pm 0.01$ & $2.10 \pm 0.01$\\ (c) &
$0.97\pm 0.02$ & $0.981 \pm 0.001$ & $0.999 \pm 0.007$ & $ 0.949 \pm
0.006$ & $1.73 \pm 0.01$\\ (d) & $0.98 \pm 0.01$ & $0.99 \pm 0.01$ &
$1.172 \pm 0.009$ & $ 0.973 \pm 0.008$ & $2.05 \pm 0.01$\\ \hline
\end{tabular}
\end{center}
\caption{The spectral tilt $n_s$ for different values of $Q$ and bulk
  viscosity parameter ${\bar \eta}_b =  \eta_b H/\rho_r$. (a) with
  radiation noise term ($C_P=0$), thermal inflaton fluctuations; (b)
  $C_P=0$, non-thermal inflaton fluctuations; (c) without the
  radiation noise term ($C_P=1$),  thermal inflaton fluctuations; (d)
  $C_P=1$, non-thermal inflaton fluctuations. These are for $N_*=50$
  and for a pivot scale of $k_0 = 1000 H_0$.}
\label{tab1}
\end{table} 

As the bulk viscosity increases, $\eta_{b*}
H_*/\rho_{r*} \gtrsim 0.18$, the power spectrum tilt quickly increases
and tends to become blue in all cases of dissipation coefficients
analyzed. This indicates that the bulk viscosity coefficient cannot be
larger than around  $\eta_{b*} H_*/\rho_{r*} \simeq 0.18$, setting,
thus an upper bound for the value of the bulk viscosity. This result
is similar to that observed only with the shear viscosity, 
$\eta_{s*} H_*/\rho_{r*} \alt 0.3$, for  $Q_* \gtrsim 10^{-4}$.

In summary, the dissipative stochastic forces we have in the
description of the relativistic fluid will always tend to enhance the
amplitude of the fluctuations in the fluid, the effect being larger on
smaller scales. The effect propagates to the field (inflaton)
fluctuations, with the corresponding enhancement of the amplitude of
the primordial spectrum. When the evolution of the background is such
that $Q$ increases during inflation, the spectrum will tend to be blue
tilted: larger wavenumbers cross the horizon at larger values of $Q_*$
for which the effect is more pronounced. Viscous effects will work in
principle in the opposite 
direction, preventing the growth of the perturbations. However in the
weak dissipative regime $Q_*< {\cal O}(1)$,  the viscous stochastic force
effect still dominates, in both the amplitude and the spectral index,
sending more power to the smaller scales. Therefore, having
a primordial spectrum with a spectral index consistent with Planck
data will constrain the amount of viscosity (bulk and shear) allowed
in the system.        
 
Before ending this section, some comments on the tensor-to-scalar
ratio $r$. In the weak dissipative regime this is given
by:
\be 
r = \frac{16 \epsilon_*}{\Delta_{Q*} + 1 + 2 {\cal N}_*} \,.  
\ee
Without dissipation, when $\Delta_{Q*}={\cal N}_*=0$, we recover the
standard cold inflation result, $r = 16 \epsilon_*$, and a value 
larger than the Planck limit $r \agt 0.3$ \cite{Planck},  being the simplest quartic
chaotic model ruled-out by observations. However dissipative dynamics,
not affecting the tensors, can render $r$ consistent with observations
due to the extra suppression factor ``$\Delta_{Q*} + 1 + 2 {\cal
  N}_*$'' \cite{Bartrum:2013fia}. 
Nevertheless, with negligible occupation number ${\cal
  N}_*\simeq 0$ and in the very weak dissipative regime, the
suppression due to $\Delta_{Q*}$ is not enough to render the ratio
consistent with observations, as was already observed in \cite{Bartrum:2013fia}. 
Besides, the value of $\epsilon_*$ is
slightly larger than in standard inflation because the value of the
field at horizon crossing is smaller, which makes $r$ increase
initially as  $Q_*$ increases . By the time the
effect of a larger $\Delta_{Q*}$ overcomes that of $\epsilon_*$, the
spectrum has become blue-tilted. However, with non-negligible
occupation number, the tensor-to-scalar ratio is always below the Planck
limit $r \alt 0.3$, with an spectral index consistent with the
data upto $Q_*\simeq 0.05$ ($Q_* \simeq 10^{-4}$) for $C_P=1$
($C_P=0$).


\section{Conclusions}
\label{sec6}

The matter content of the very early universe generically consists
of a multi-particle system with a wide range of particle properties
and interactions.  
Neglecting some of this richness can lead to missing out some
important physical phenomena. Cold inflation is an idealization where 
the dynamics reduces to the classical evolution of the scalar inflaton field
with vacuum quantum fluctuations superposed on this background field.
Warm inflation includes additional multiparticle dynamics and recent
success of its  
predictions \cite{Bartrum:2013fia}
in fitting the Planck results provides support that these effects may have an
important role to play.  In this, and other 
situations where radiation is present
in the early Universe, the idealization is often made of a perfect
fluid, whereas there might be some deviations from this limit
that lead to viscous dissipation and corresponding noise forces, and these
effects might have observational consequences.  
One example where this could be applied is to the many studies looking
at thermal fluctuations seeding density perturbations in a
radiation dominated regime 
\cite{Magueijo:2002pg,hmb,Bhattacharya:2005wn,Lieu:2011rj,Biswas:2013lna}.
To provide a framework in which all these types of problems can be
examined, this paper has obtained the coupled set of equations of a
scalar field with dissipation interacting with an imperfect radiation fluid
and treating also the corresponding density perturbations.

As an example, we have applied the equations to the warm inflation
scenario, and study how the different stochastic forces affects the
primordial spectrum. Einstein equations do not fully fix whether the
dissipative noise source enters in the energy flux ($C_P=0$) or in the momentum
flux ($C_P=1$), and we have explored and compared both possibilities. 
Previous studies of the primordial spectrum in warm inflation only took into
account the second possibility with $C_P=1$
\cite{mossgraham,shear,Ramos:2013nsa}. It was shown that for a $T$
dependent dissipative coefficient the amplitude of the spectrum gets
enhanced for values of the inflaton dissipation term larger than the
Hubble parameter, $Q_* \agt 1$. The same behavior is obviously
present when the noise sources directly the radiation energy density. 
But before the growing mode dominates the behavior of the fluctuations, the
stochastic source will increase further the amplitude. 
 In a model like the quartic potential considered here, for
 which $Q$ increases during inflation,  the effect is larger on
 smaller scales and the tilt of the spectrum increases.
 Nevertheless, for low enough values of  $Q_* \alt 10^{-4}$ the 
 effect is negligible and the spectral index remains within Planck
 limits. However, in order to obtain a tensor-to-scalar ratio also
 within the Planck upper bound, we need to consider a non-trivial (thermal)
 statistical distribution of inflaton fluctuations
 \cite{Bartrum:2013fia}.  

We have shown that the viscosity terms act to strongly damp the
radiation perturbations in the regime where the dissipation of the
inflaton field is large (compared to the Hubble parameter). Thus, the
viscosities tend to counter balance the effect of the growing mode
observed in ~\cite{mossgraham}.  However, because of the random
noise terms in the radiation fluid equations, the growing mode is not
completely eliminated. In fact, we also have shown that the random
noise caused by the viscous radiation fluid tends to further
contribute to the  curvature perturbation spectrum causing it to
increase even in the low dissipative regime of warm inflation
(when the inflation dissipation term is smaller than the Hubble
parameter).   Since the random fluctuation contributions that are added
to the curvature perturbation spectrum are proportional to the
viscosity coefficients, for models where viscosity increases during
inflation this implies more power at smaller scales, i.e., a larger
tilt. And this allows us to put strong constraints on the
level of viscosity permissible in the warm inflation scenario.


\acknowledgments

M.BG. is partially supported by MICINN (FIS2010-17395) and "Junta de Andalucia"
(FQM101).
A.B. is partially supported by a UK Science and Technology Facilities Council
Consolidated Grant.
I.G.M. is partially supported by 
the UK Science and Technology Facilities Council Consolidated 
Grant ST/J000426/1.
R.O.R. is partially supported by research grants from the brazilian
agencies  Conselho Nacional de Desenvolvimento Cient\'{\i}fico e
Tecnol\'ogico (CNPq) and  Funda\c{c}\~ao Carlos Chagas Filho de Amparo
\`a Pesquisa do Estado do Rio de Janeiro (FAPERJ).


\end{document}